\documentclass[twocolumn,aps,showpacs,nofootinbib,prd]{revtex4-1}
\usepackage{amssymb}
\usepackage{amsmath}
\usepackage{graphics}
\usepackage{epsfig}
\usepackage[dvips]{color}

\newcommand{\be}{\begin{equation}}
\newcommand{\ee}{\end{equation}}
\newcommand{\bea}{\begin{eqnarray}}
\newcommand{\eea}{\end{eqnarray}}
\newcommand{\nn}{\nonumber}
\newcommand{\beas}{\begin{eqnarray*}}
\newcommand{\eeas}{\end{eqnarray*}}

\newcommand{\pa}[1]{\left( #1 \right) }
\newcommand{\co}[1]{\left[ #1 \right] }

\newcommand{\bd}[1]{{\bf #1}}

\begin{document}

\title{Head shock vs Mach cone: \\
azimuthal correlations from 2$\rightarrow$3 parton processes \\
in relativistic heavy-ion collisions} 
\author{Alejandro Ayala$^1$, Isabel Dominguez$^1$ and Maria Elena
  Tejeda-Yeomans$^2$} \address{$^1$Instituto de Ciencias
  Nucleares, Universidad Nacional Aut\'onoma de M\'exico, Apartado
  Postal 70-543, M\'exico Distrito Federal 04510,
  Mexico.\\ $^2$Facultad de Ciencias F\'isico-Matem\'aticas, Universidad Aut\'onoma de Sinaloa,
Avenida de las Am\'ericas y Boulevard Universitarios, Ciudad Universitaria,
C.P. 80000, Culiac\'an, Sinaloa, M\'exico.\\ $^3$ Departamento de F\'{\i}sica,
  Universidad de Sonora, Boulevard Luis Encinas J. y Rosales, Colonia
  Centro, Hermosillo, Sonora 83000, Mexico.}

\begin{abstract}

We study the energy momentum deposited by fast moving partons within a
medium using linearized viscous hydrodynamics. The particle
distribution produced by this energy momentum is computed using the
Cooper-Frye formalism. We show that for the conditions arising in
heavy-ion collisions, energy momentum is preferentially deposited
along the head shock of the fast moving partons. We also show that the
double hump in the away side of azimuthal correlations can be produced
by two (instead of one) away side partons that deposit their
energy momentum along their directions of motion. These partons are
originated in the in-medium hard scattering in $2\to 3$ processes. We
compare the results of the analysis to azimuthal angular correlations
from PHENIX and show that the calculation reproduces the data
systematics of a decreasing away side correlation when the momentum of
the associated hadron becomes closer to the momentum of the leading
hadron. This scenario seems to avoid the shortcomings of the Mach cone
as the origin of the double-hump structure in the away side.

\end{abstract}

\maketitle

\section{Introduction}\label{I}

Azimuthal angular correlations have provided a powerful testing ground
to elucidate the propagation properties of fast partons within the
medium created in high-energy heavy-ion reactions~\cite{STAR,PHENIX}.
The main features of these correlations can be summarized as follows:
when the leading and the away side particles have similar momenta,
the correlation shows a suppression of the away side peak, compared to
proton collisions at the same energies. However, when the momentum
difference between leading and away side particles increases, either a
double peak or a broadening of the away side peak appears, whereas
neither of these are present in proton collisions at the same
energies~\cite{azcor}. The peak region is known as the shoulder and
the region between the peaks is known as the head. For low-momentum
particles, the jet multiplicity is larger in the shoulder than in the
head region~\cite{PHENIX}.

Because the structures in the away side are best seen for low momentum
particles, there were explanations based on the emission of sound
modes caused by one fast moving parton~\cite{Casalderrey,Renk}, the
so-called {\it Mach cones}. Nevertheless, it has been argued that such
interpretation is fragile, since the jet-medium interaction produces
also a wake whose contribution cannot be
ignored~\cite{Casalderrey2,Torrieri}. Furthermore, it was recently
shown that it is unlikely that the propagation of one high-energy
particle through the medium leads to a double-peak structure in the
azimuthal correlation in a system of the size and finite viscosity
relevant for heavy-ion collisions, since the energy momentum
deposition in the head shock region is strongly
forwardpeaked~\cite{Bouras}. Moreover, by using a realistic
multiple-gluon emission for the parton shower produced by an in-medium
moving parton, the overlapping perturbations in very different spatial
directions wipe out any distinct Mach cone structure ~\cite{Neufeld3}.

The more widely accepted explanation for the double peak or broadening
of the away side is given nowadays in terms of initial state
fluctuations of the matter density in the colliding nuclei. These
fluctuations are then shown to give rise to an anisotropic flow of
partially equilibrated, low-momentum particles, within the bulk
medium. Hydrodynamic descriptions of this scenario have for instance
successfully reproduced the experimental $v_3$~\cite{v3}.  However, it
has also been shown that there is a strong connection between the
observed away side structures and the medium's path
length~\cite{pathlength}.  The connection is expressed through the
dependence of the azimuthal correlation on the trigger particle
direction with respect to the event plane in such a way that, for
selected trigger and associated particle momenta, the double peak is
present (absent) for out-of-plane (in-plane) trigger particle direction.
A final-state effect rather than an an initial state one seems more
consistent with this observation.

\begin{figure*}[t] 
{\centering
  \hspace{-1cm}
  {\includegraphics[scale=0.6]{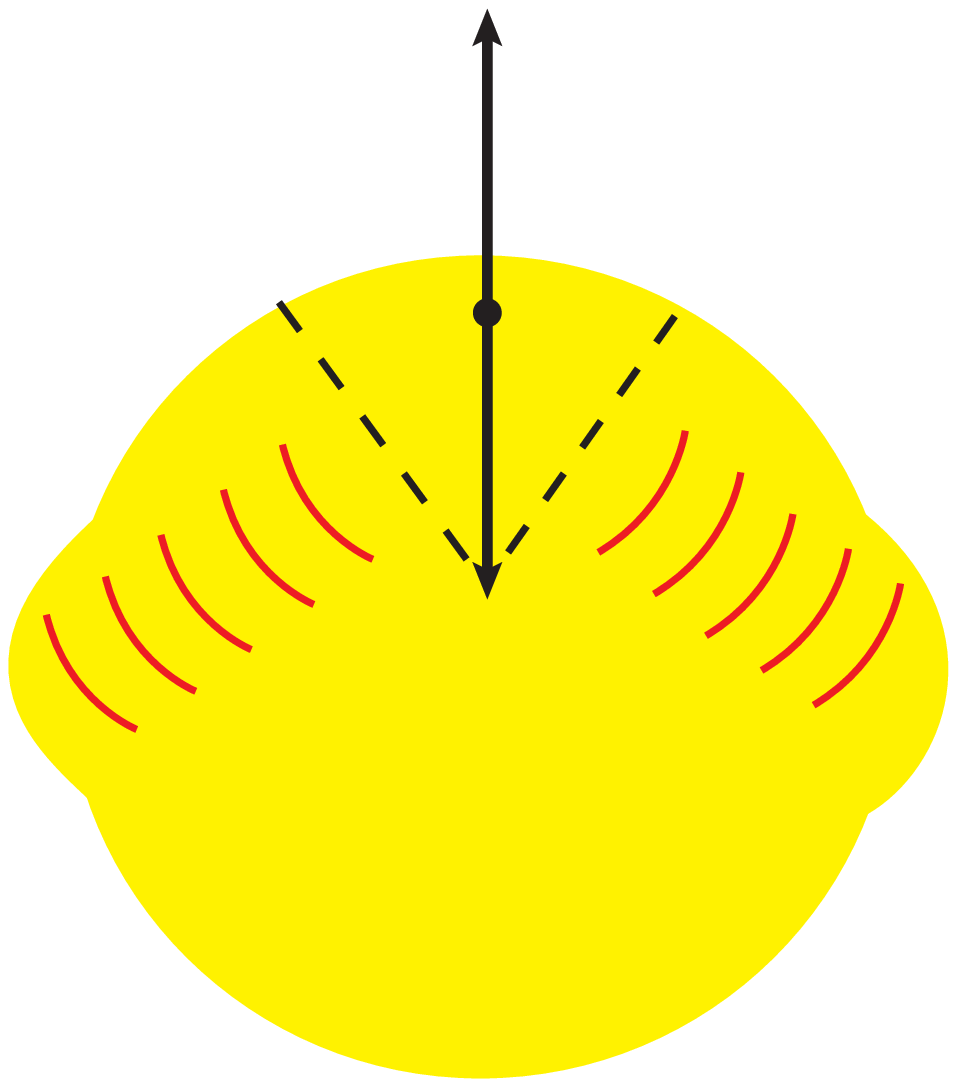}}
  \hspace{1cm}
  {\includegraphics[scale=0.6]{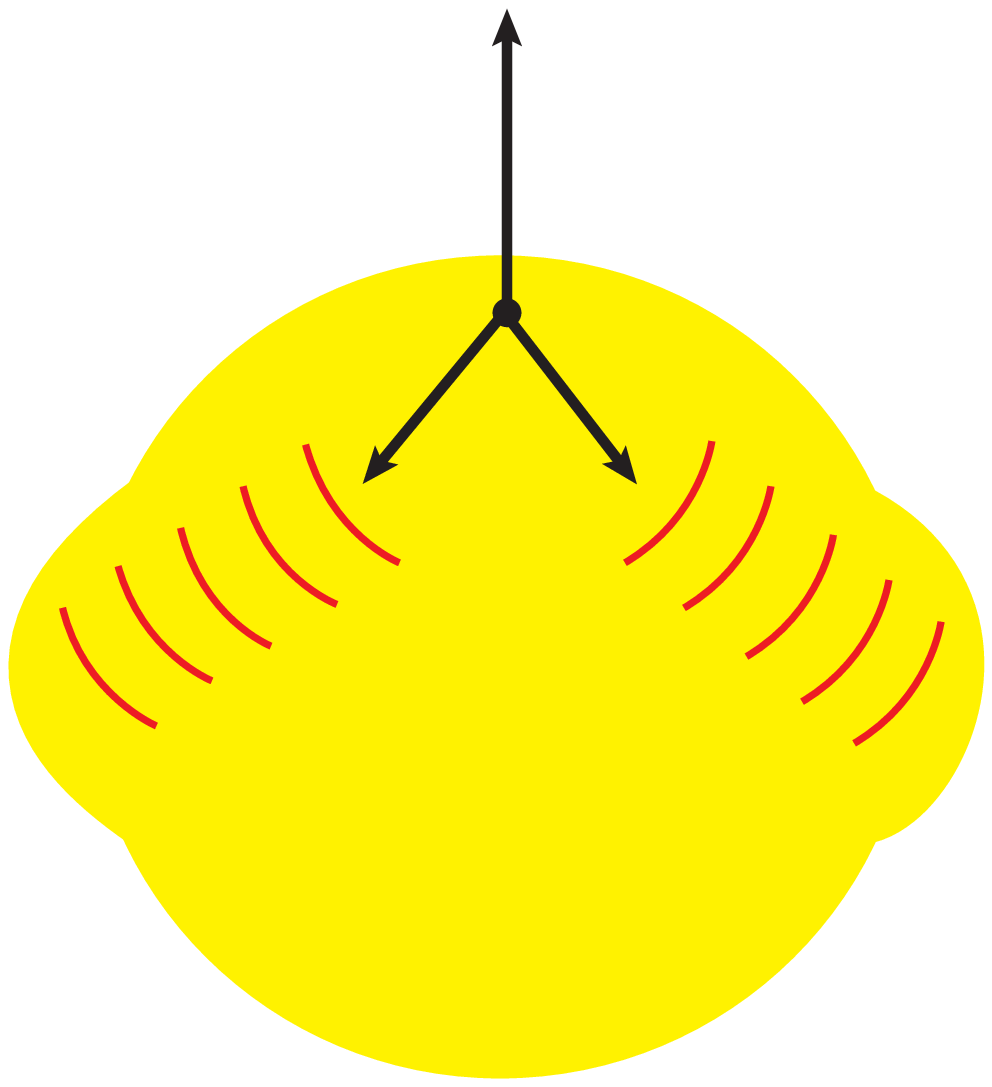}
}}
\caption{(Color on line) Schematic representation of the possible
  origin of the away side double hump. The left panel illustrates the
  common picture whereby one away side parton deposits
  energy momentum, exciting the sound modes and producing a Mach
  cone. The right panel illustrates our approach whereby two away side
  partons each generate head-shock and deposit energy momentum
  preferentially along their directions of motion.}
\label{fig1}
\end{figure*}

The medium's response to a fast moving parton can be described by the
energy and momentum that the parton deposits. This energy momentum is
then converted into particles upon hadronization. When the interaction
produces low-momentum particles, a hydrodynamic description of the
medium's response seems appropriate.  In this framework, it is natural
to explore the types of medium excitations generated by fast moving
partons and the way these in turn are responsible for the shape of the
away side in the angular correlations. When the \textit{sound} (that
is, the \textit{longitudinal modes}) are prominently excited, the
double peak in the away side could emerge as a Mach cone structure,
that is, as energy momentum mainly deposited on the sides of the path
traveled by the parton. This is illustrated in the left-hand panel of
Fig.~\ref{fig1}. However, when the modes that get prominently excited
are the \textit{wake} (that is, the \textit{transverse modes}), energy
and momentum is preferentially deposited along the direction of the
travelling parton. If a double peak is to emerge, the alternative
scenario is that the structure is produced by two travelling partons
in the away side. This is illustrated in the right-hand panel of
Fig.~\ref{fig1}.

\begin{figure}[b] 
\begin{center}
  \includegraphics[scale=0.4]{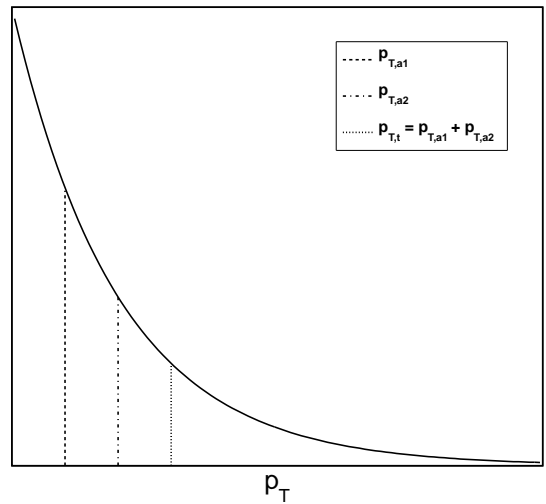}
  \caption{Schematic representation of how a single away side parton is
  less abundantly produced than two away side partons carrying the
  same total momentum, given that the momentum distribution function
  rapidly decays with momentum.}
\label{fig2}
\end{center}
\end{figure}

Two partons in the away side are produced in processes where two
initial-state partons scatter into three final-state ones, the
so-called $2\rightarrow 3$ processes. In this case the cross section
is smaller than the one for $2\rightarrow 2$ processes, at least by
one power of $\alpha_s$. However, consider a scattering event that
deposits a given amount of energy in the away side. If such an event
comes from a $2\rightarrow 2$ process, the away side parton will have
a larger energy than each of the two partons, when these last come
from a $2\rightarrow 3$ process. Because the parton distribution is a
fast falling function of the parton energy, the extra power of
$\alpha_s$ is partially compensated by the larger abundance of partons
with smaller energy. This is illustrated in Fig.~\ref{fig2}. For
$2\rightarrow 3$ processes, when the dominant energy momentum
deposition happens through transverse modes, conservation of momentum
at the parton level gives rise to a distinctive angular dependence in
the azimuthal correlation whereby, the angular difference between the
peaks in the away side is close to $2\pi/3$ rad.

The computation of hadron events produced by $2\rightarrow 3$ parton
processes in the context of azimuthal correlation functions was put
forward and explored in Refs.~\cite{Ayala1, Ayala2}, using the leading
order QCD matrix elements. Those studies were made for fragmentation
outside the medium. In this work we compute the hadron multiplicity
produced by two partons in the away side coming from $2\rightarrow 3$
processes for the case where all its energy momentum is deposited into
the medium. We use linearized viscous hydrodynamics to compute the
medium's response. The work is organized as follows: In Sec.~\ref{II}
we first review the general framework to compute the particle
distribution stemming from a given energy momentum deposited by a fast
moving parton within the medium, using linearized viscous
hydrodynamics. This energy momentum is then converted into a parton
multiplicity using the Cooper-Frye formalism~\cite{Cooper-Frye}.  We
discuss the different shapes for the azimuthal correlations that are
obtained when varying the strength of the wake and sound mode
contributions. In Sec.~\ref{III} we solve the hydrodynamic equations
for the transverse and longitudinal modes. We work in the limit where
the parton's velocity is larger than the sound velocity. In
Sec.~\ref{IV} we use these solutions to compute the azimuthal
correlations by convoluting the pQCD probability to produce two away
side partons in $2\rightarrow 3$ processes with the multiplicity
obtained from the excess of energy momentum produced by the moving
partons followed by fragmentation.  We finally summarize and conclude
in Sec.~\ref{V}.

\section{Particle distribution}\label{II}

We compute the particle's multiplicity as given by the Cooper-Frye formula
\bea
   E\frac{dN}{d^3p}=\frac{1}{(2\pi )^3}\int d\Sigma_\mu p^\mu
   [f(p\cdot u) - f(p_0)],
   \label{Cooper-Frye}
\eea 
where $f(p\cdot u) - f(p_0)$ is the phase-space disturbance
produced by the fast moving parton on top of the equilibrium
distribution $f(p_0)$, with $\Sigma_\mu$ and $p_\mu$ representing the
freeze-out hypersurface and the particle's momentum,
respectively. The medium's total four-velocity $u^\mu \equiv u_0^\mu +
\delta u^\mu$ is made out of two parts: the background four-velocity
$u_0^\mu$ and the disturbance $\delta u^\mu$. This last contribution
is produced by the fast moving parton and can be computed using
viscous linear hydrodynamics once the source, representing the parton,
is specified. For a static background (which hereby we assume) and in
the linear approximation, $u^\mu$ can be written as \bea u^\mu
&\equiv& u_0^\mu + \delta u^\mu
\nn\\ &=&\left(1,\frac{\mathbf{g}}{\epsilon_0(1+c_s^2)}\right),
   \label{fourvel}
\eea
where the spatial part of the medium's four-velocity,
${\mathbf{u}}={\mathbf{g}}/\epsilon_0(1+c_s^2)$, is written for
convenience in terms of the momentum density ${\mathbf{g}}$ associated
to the disturbance, with $\epsilon_0$ and $c_s$ the static
background's energy density and sound velocity, respectively. We focus
on events at central rapidity, $y\simeq 0$, and take the direction of
motion of the fast parton to be the $\hat{z}$ axis and the beam axis
to be the $\hat{x}$ axis. With this geometry, the transverse plane is
the $\hat{y}-\hat{z}$ plane and therefore, the momentum four-vector
for a (massless) particle is explicitly given by \bea p_\mu &=&
(E,p_x,p_y,p_z)\nn\\ &=& (p_T,0,p_T\sin\phi,p_T\cos\phi),
   \label{momcomponents}
\eea
\begin{figure}[t] 
\begin{center}
  \includegraphics[scale=0.55]{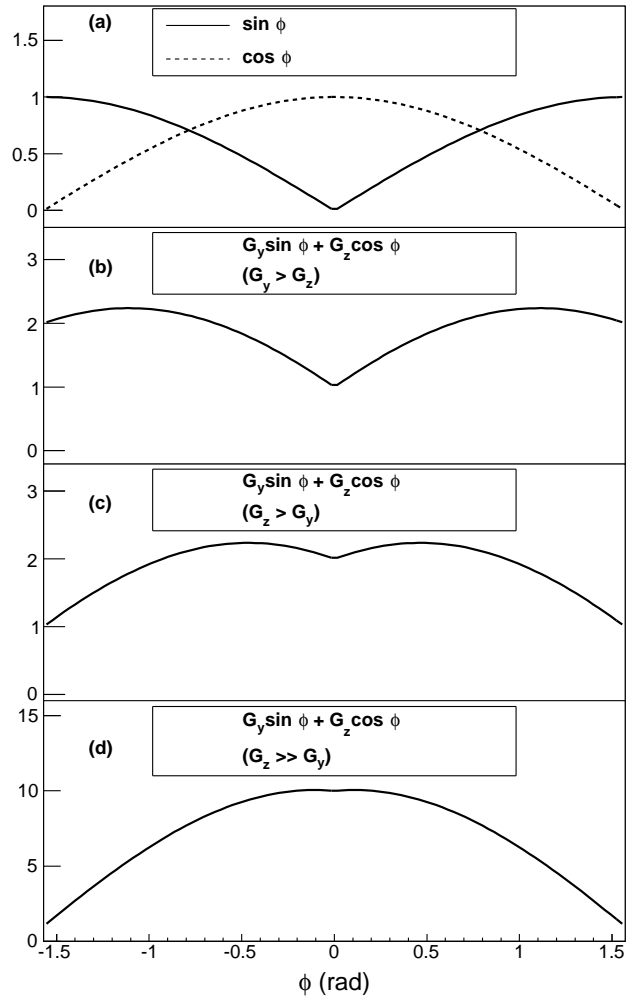}
\caption{Schematic representation of the particle distribution,
  arising from the Cooper-Frye formula. The distribution is centered
  around the direction of motion ($\phi=0$) of the parton that
  deposits the energy momentum in the medium. Since the distribution
  is a sum of a $\sin(\phi)$ and a $\cos(\phi)$, its shape depends on
  the relative strength of their respective coefficients, $G_y$ and
  $G_z$. We show first the (a) $\sin(\phi)$ and $\cos(\phi)$
  functions, (b) the case where $G_y>G_z$, (c) the case where
  $G_z>G_y$, and finally (d) the case where $G_z\gg G_y$.}
\label{fig3}
\end{center}
\end{figure}

\begin{figure*}[t] 
\begin{center}
\includegraphics[scale=0.25]{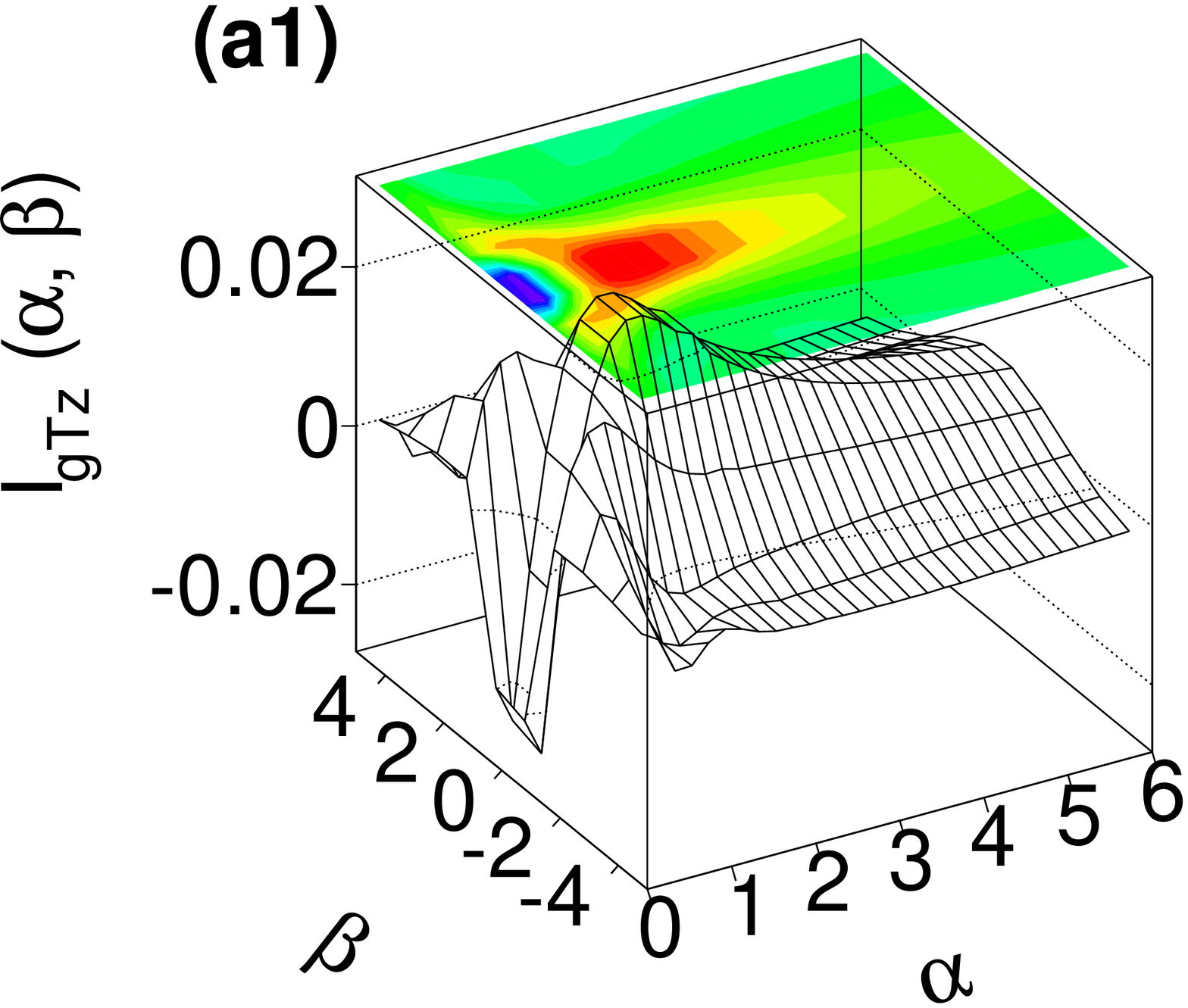}\includegraphics[scale=0.25]{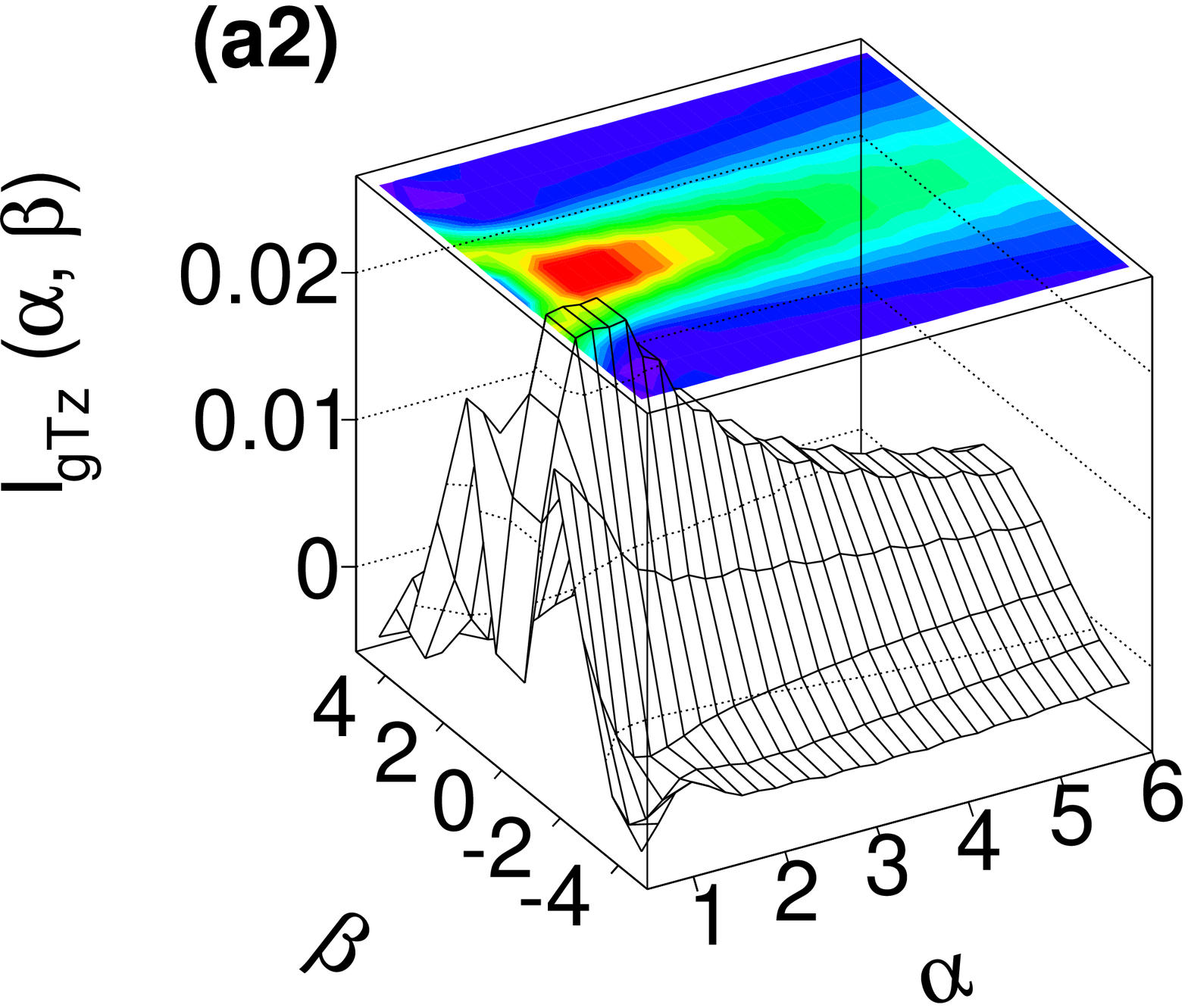}\includegraphics[scale=0.25]{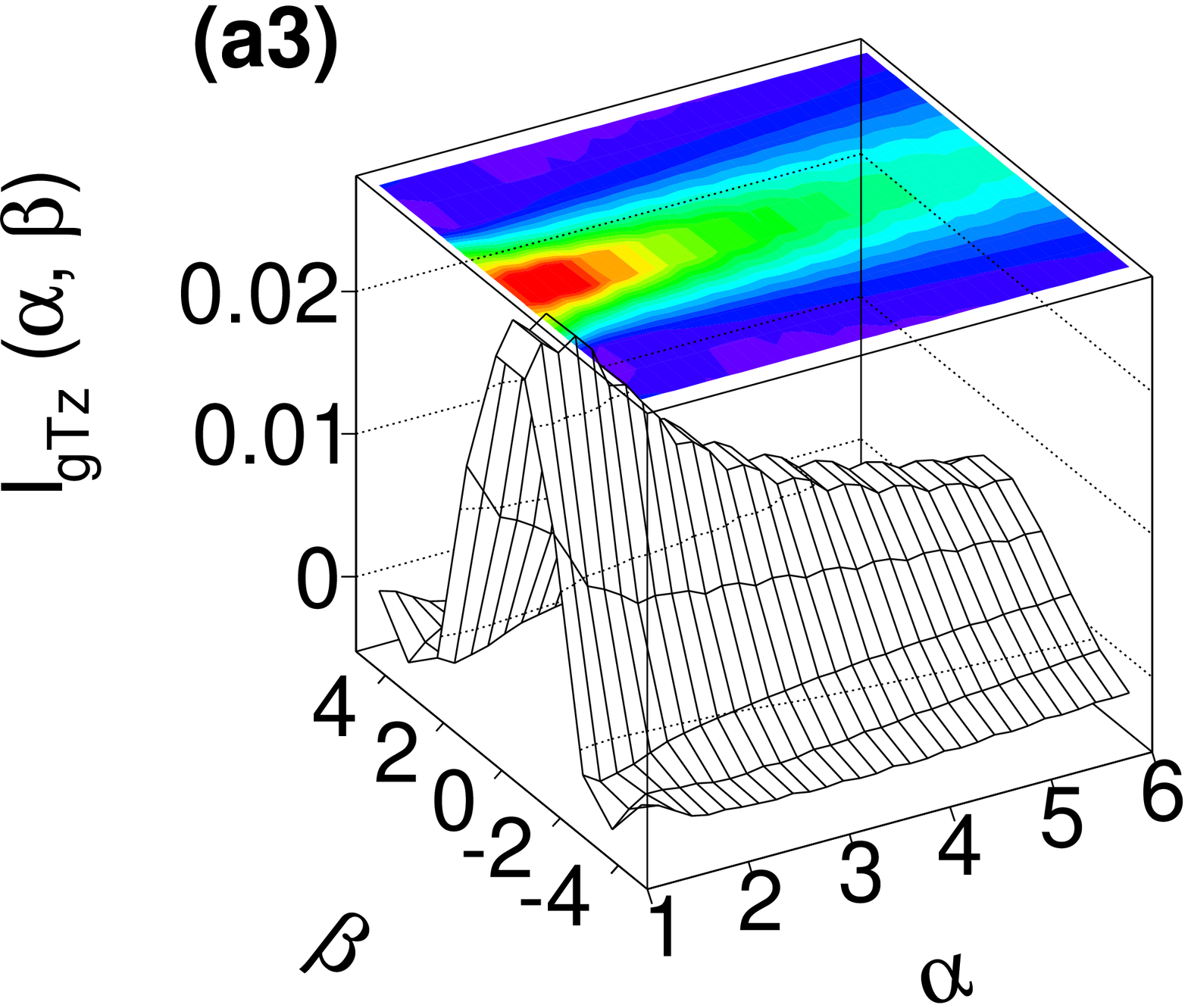}
\includegraphics[scale=0.25]{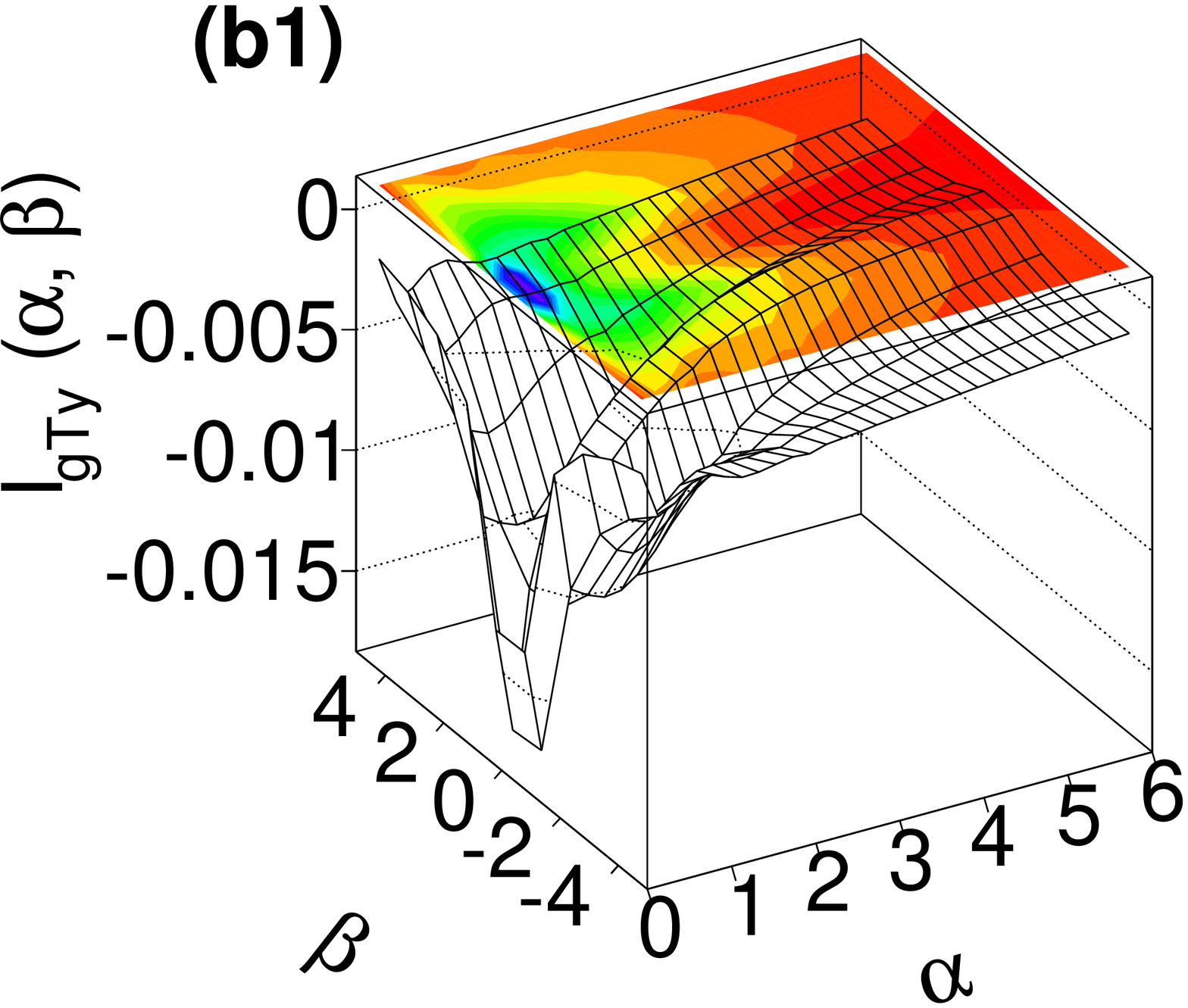}\includegraphics[scale=0.25]{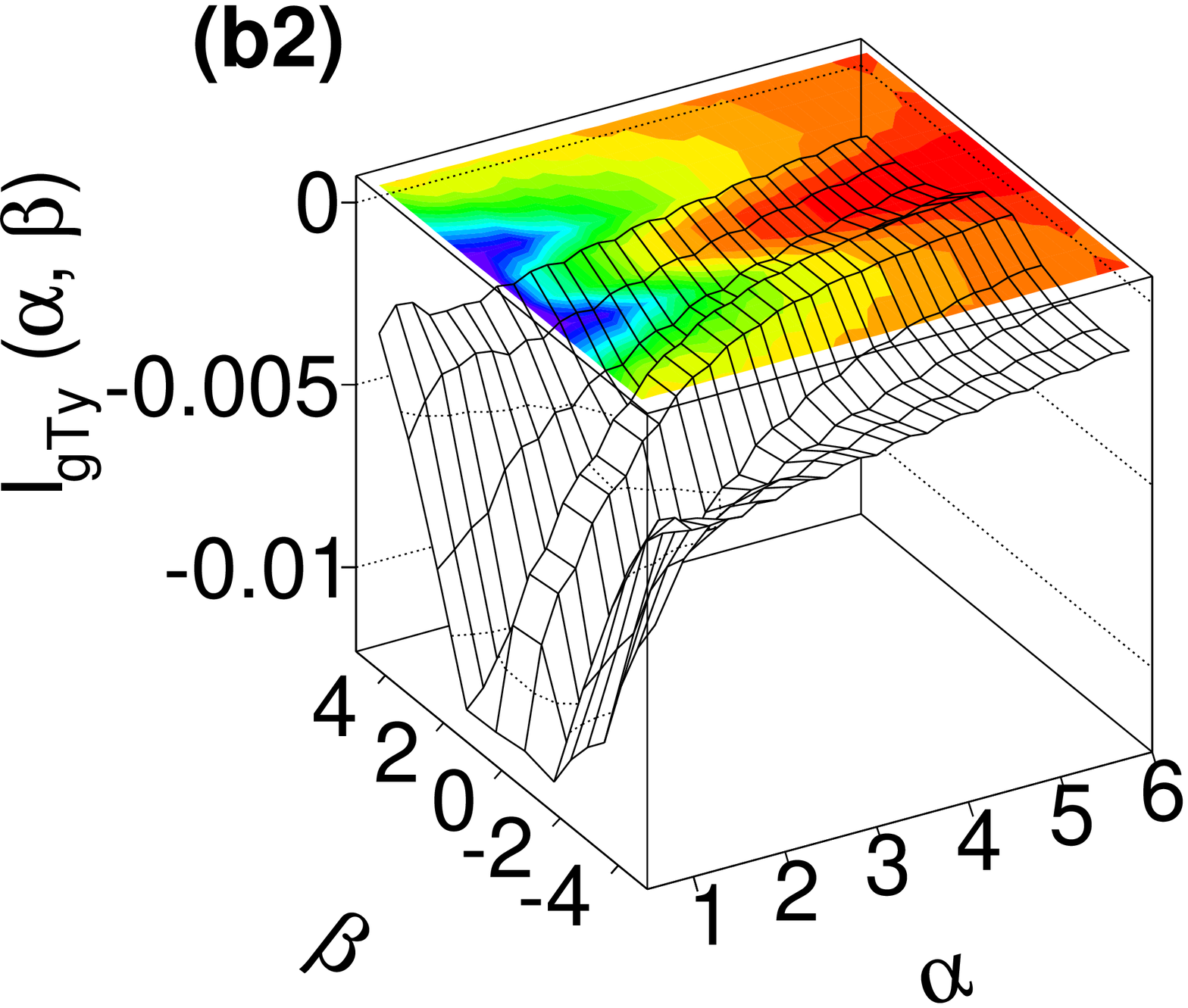}\includegraphics[scale=0.25]{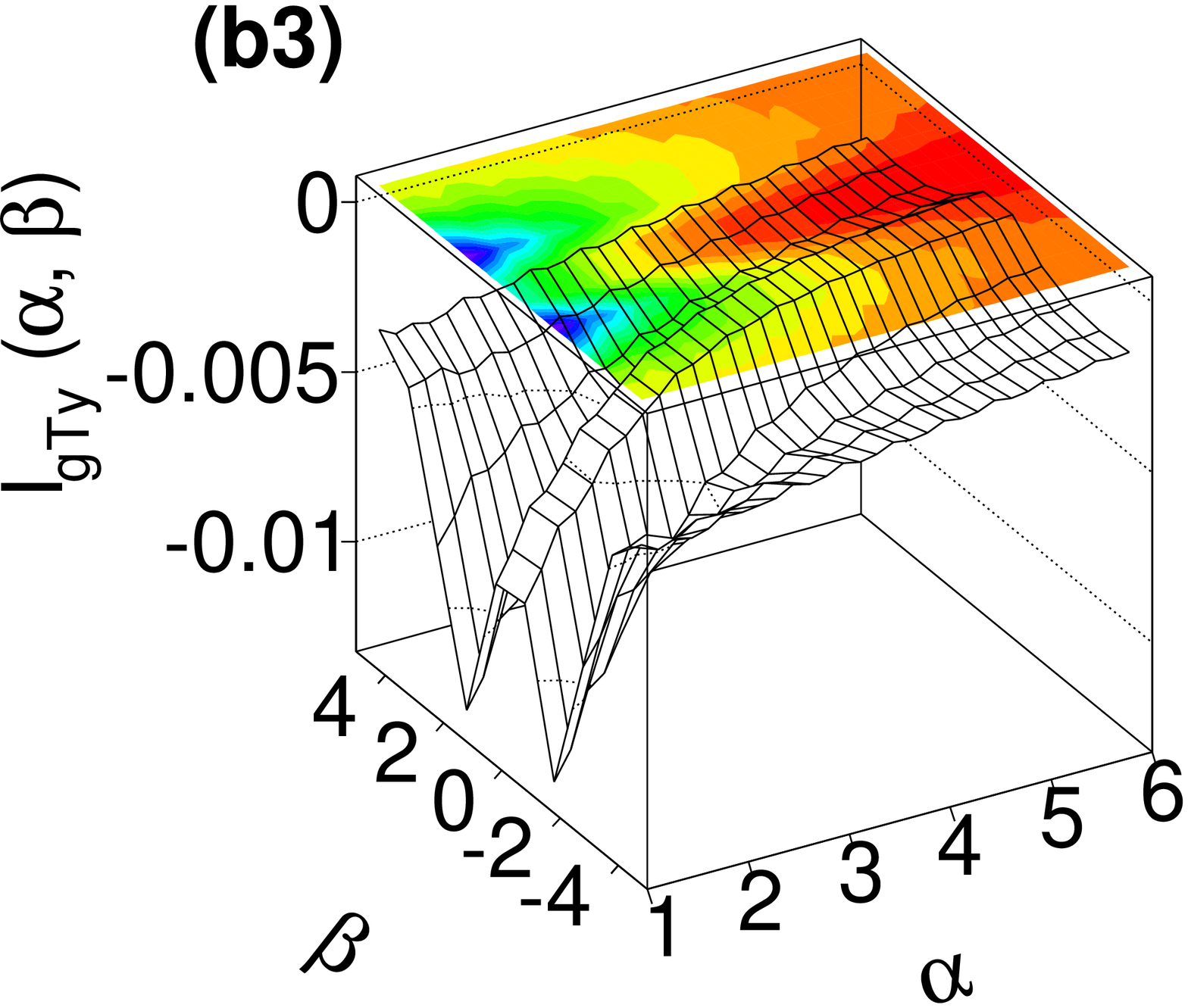}
\includegraphics[scale=0.25]{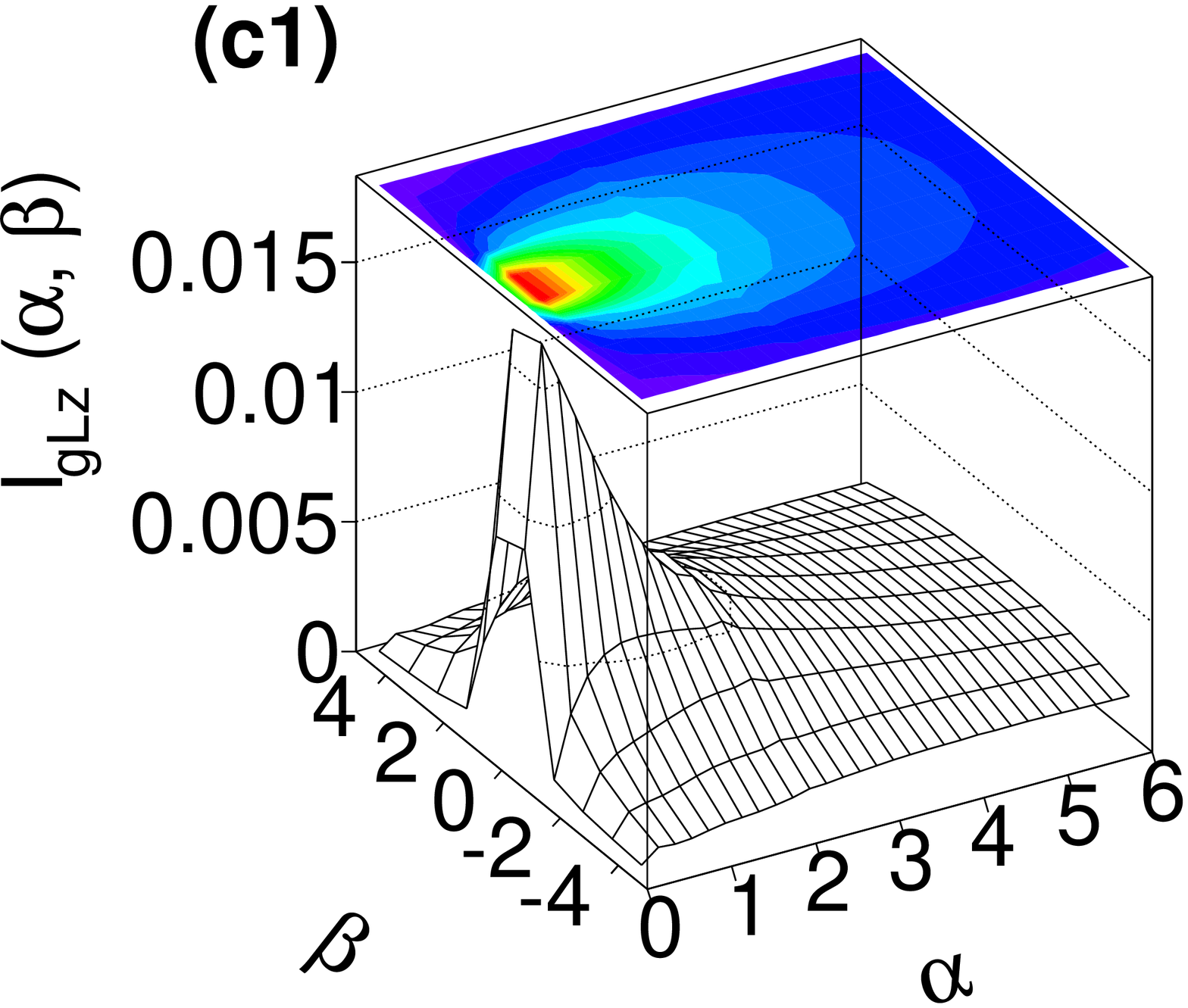}\includegraphics[scale=0.25]{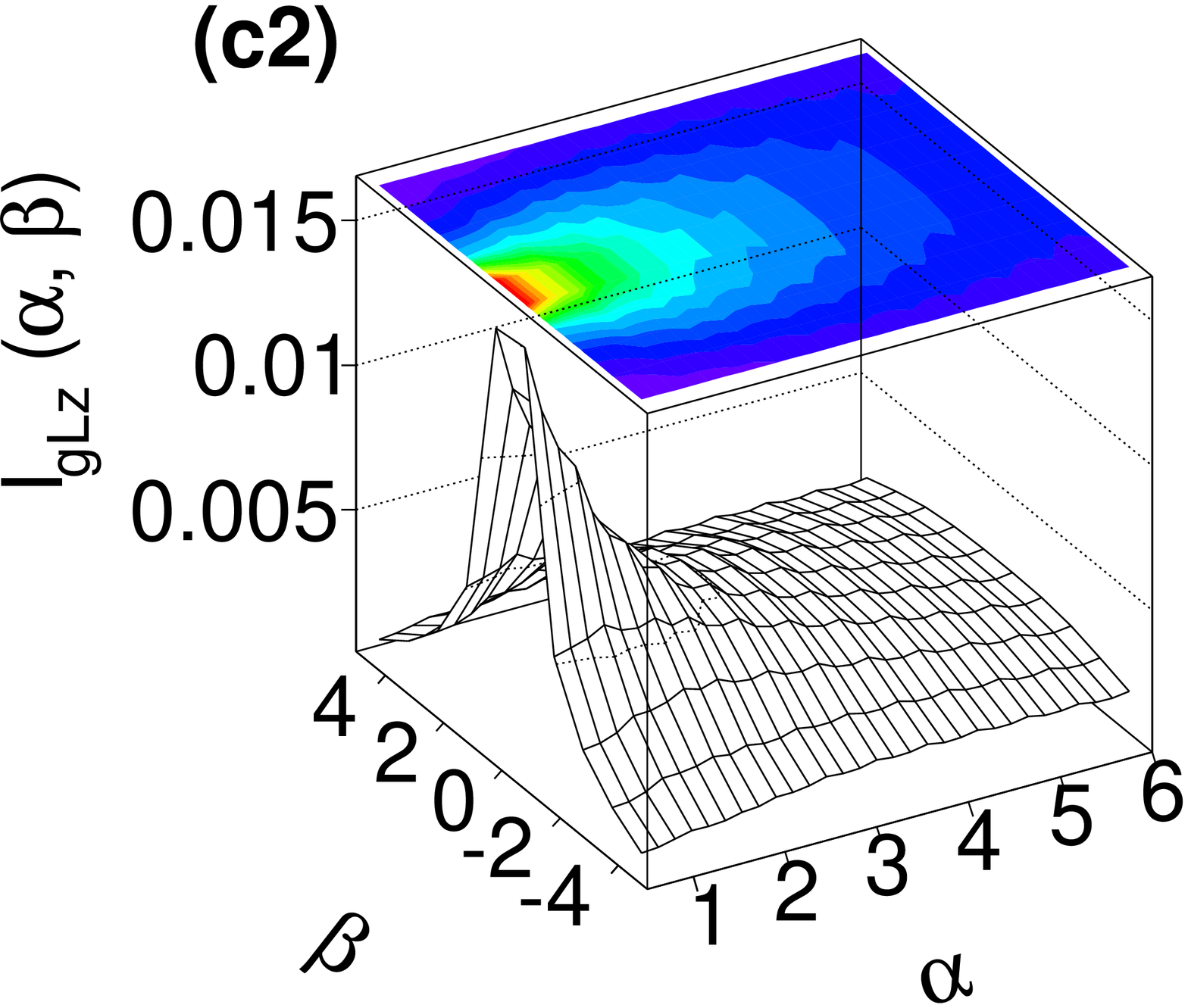}\includegraphics[scale=0.25]{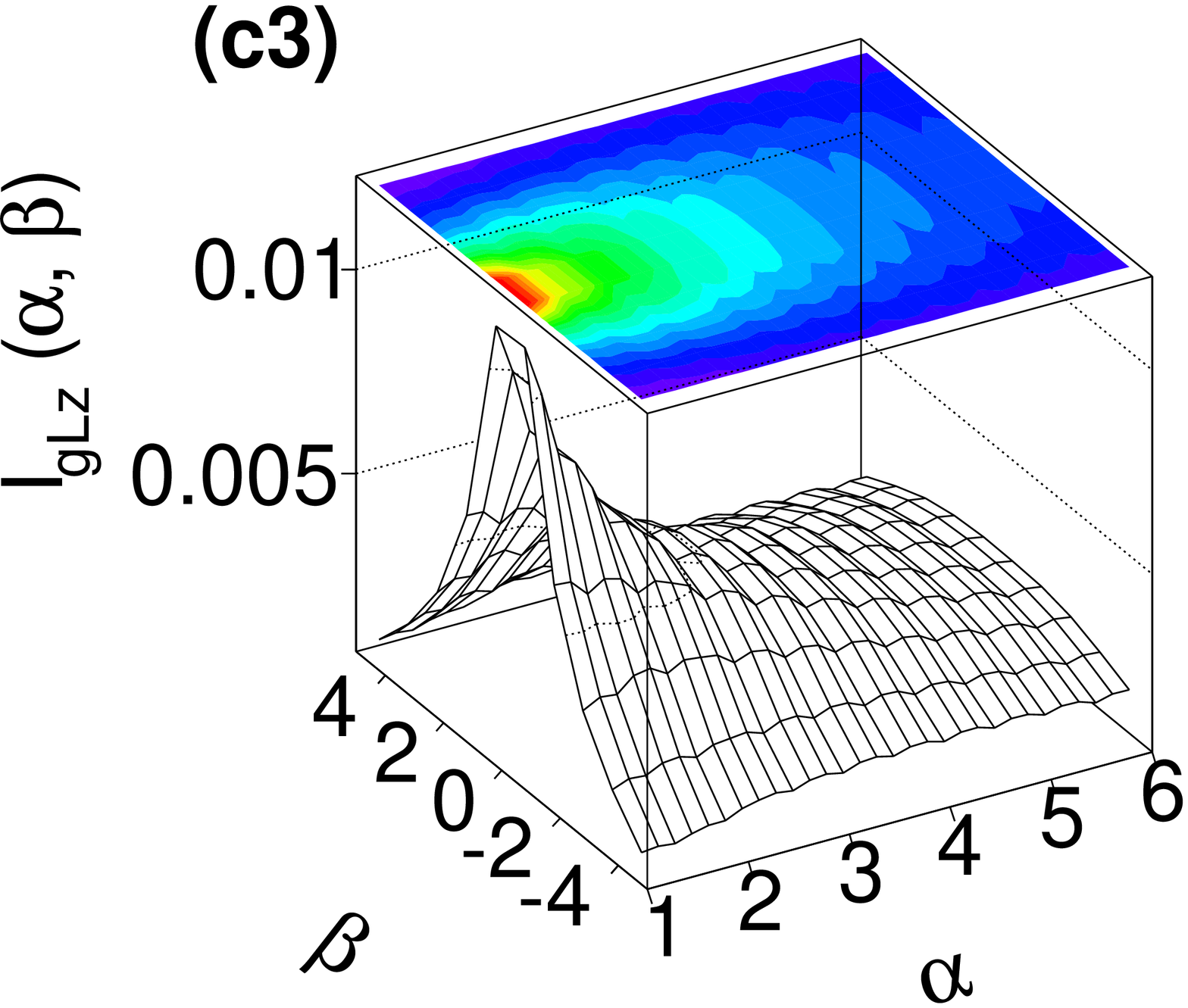}
\includegraphics[scale=0.25]{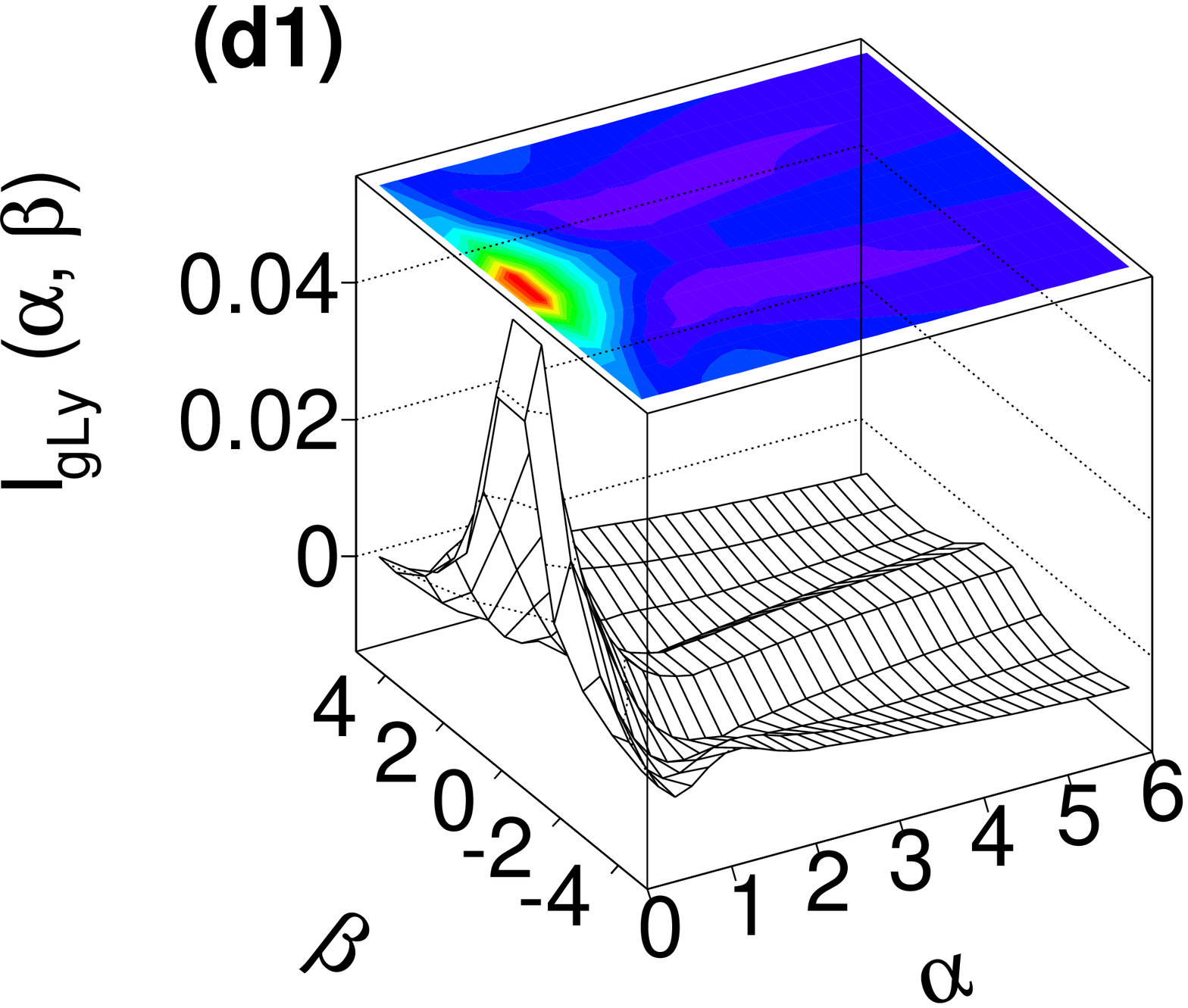}\includegraphics[scale=0.25]{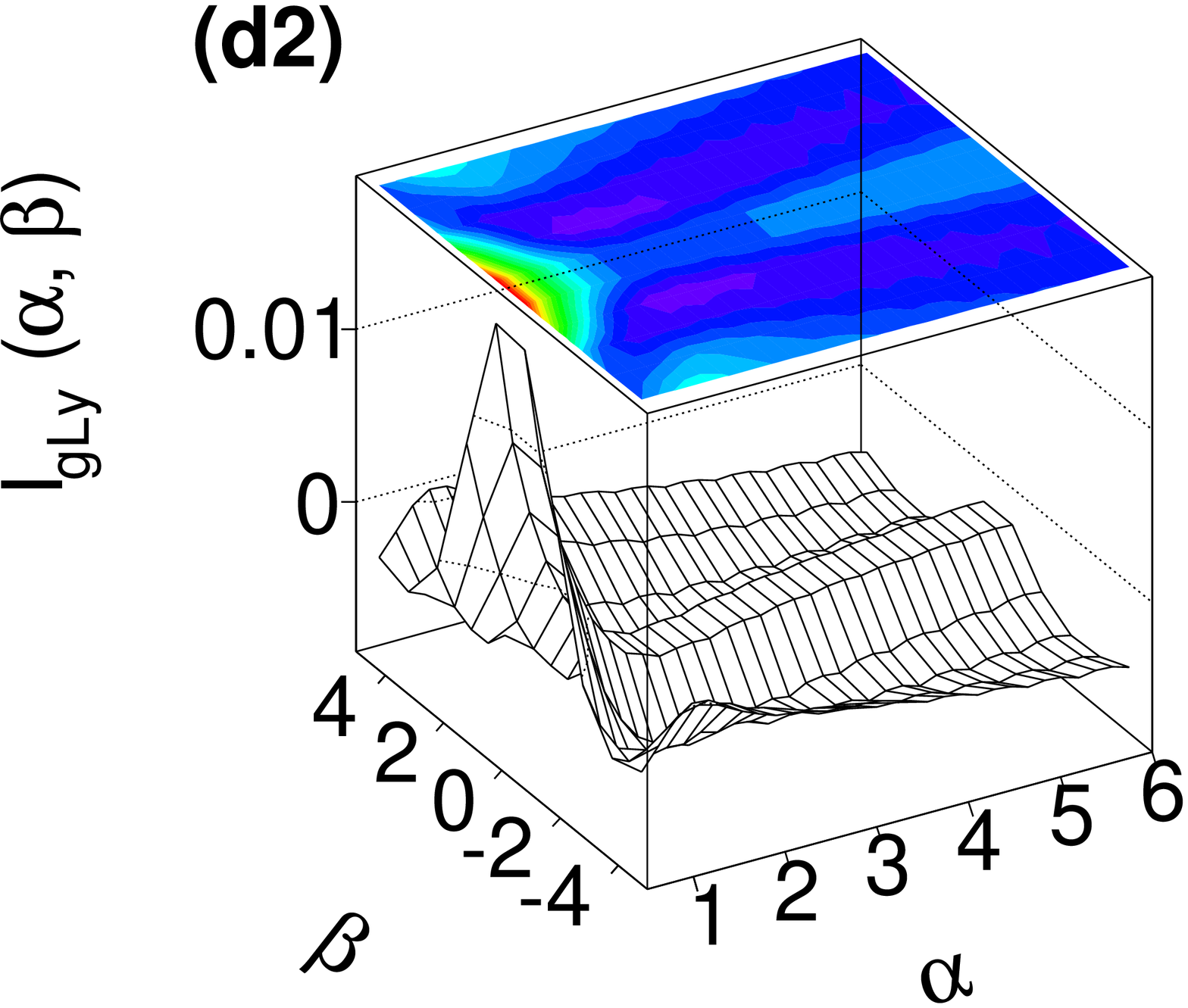}\includegraphics[scale=0.25]{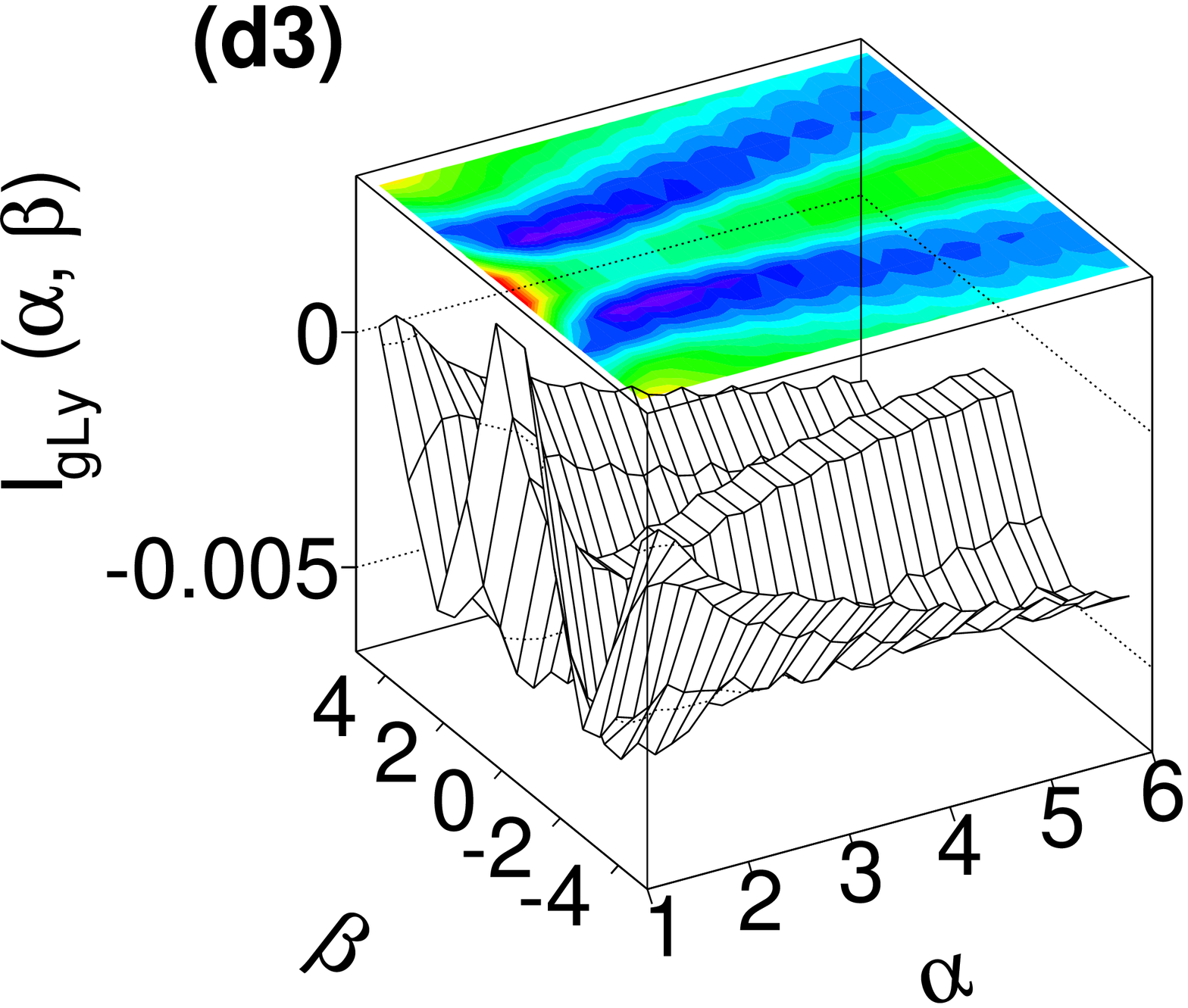}
\includegraphics[scale=0.25]{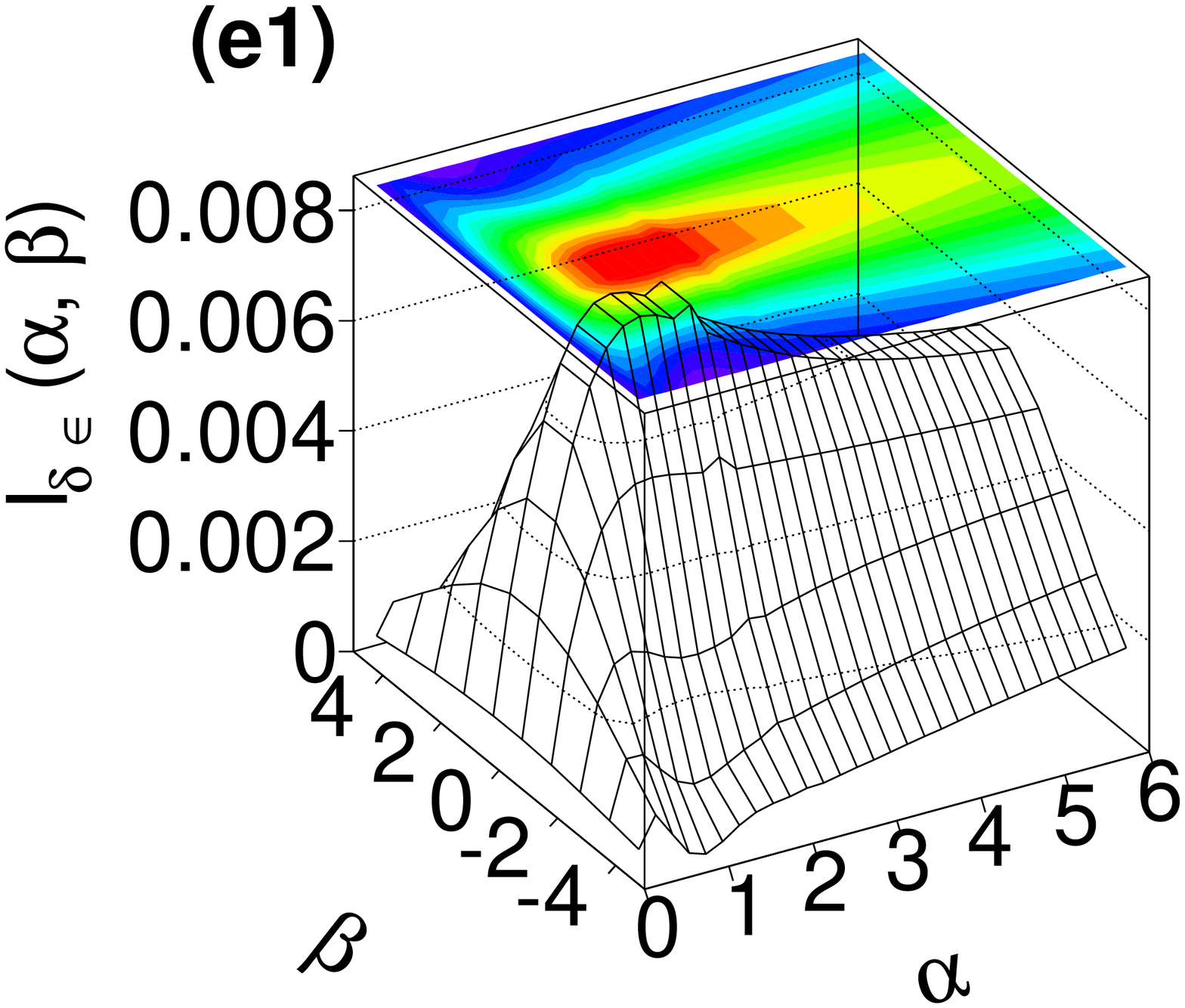}\includegraphics[scale=0.25]{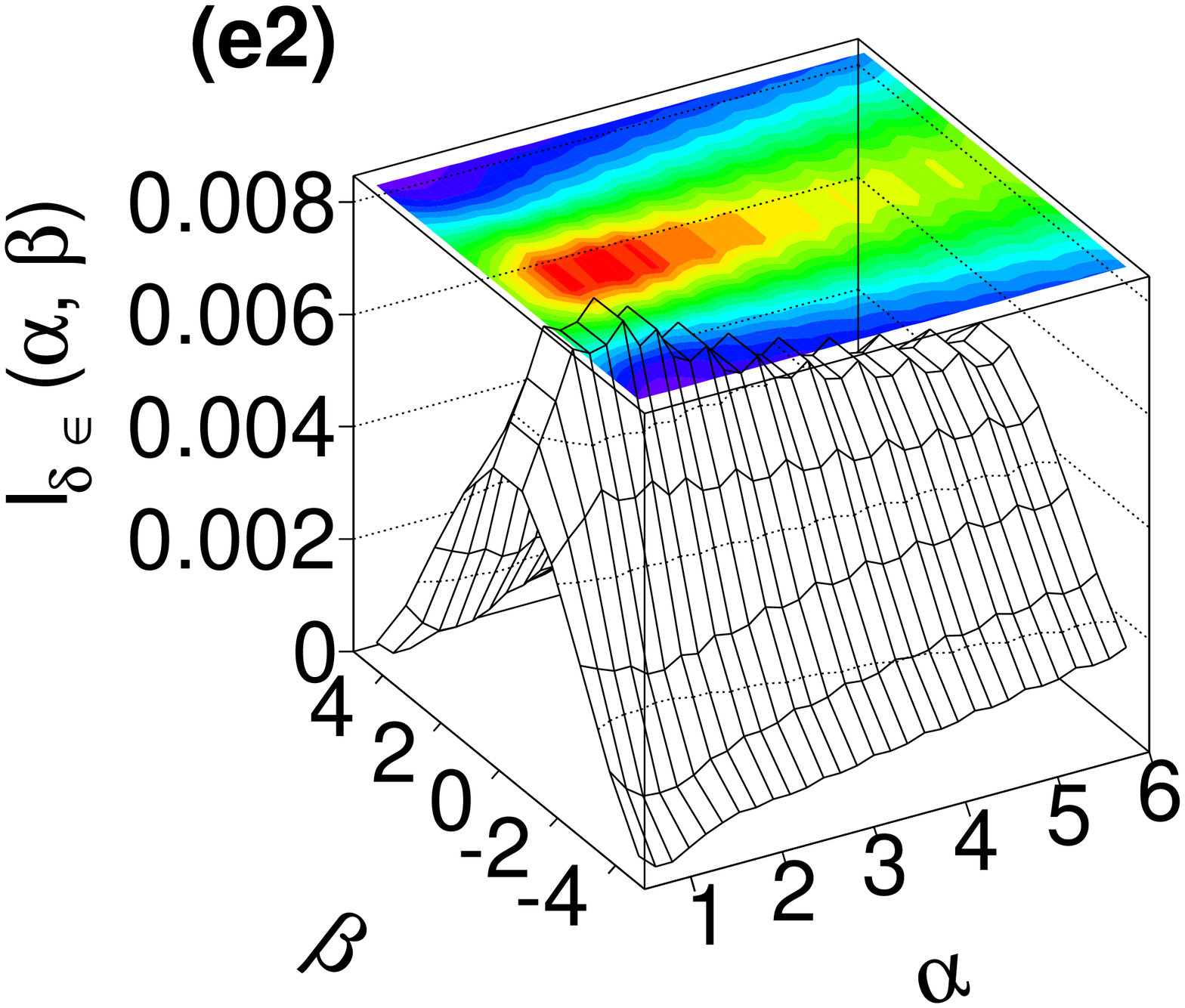}\includegraphics[scale=0.25]{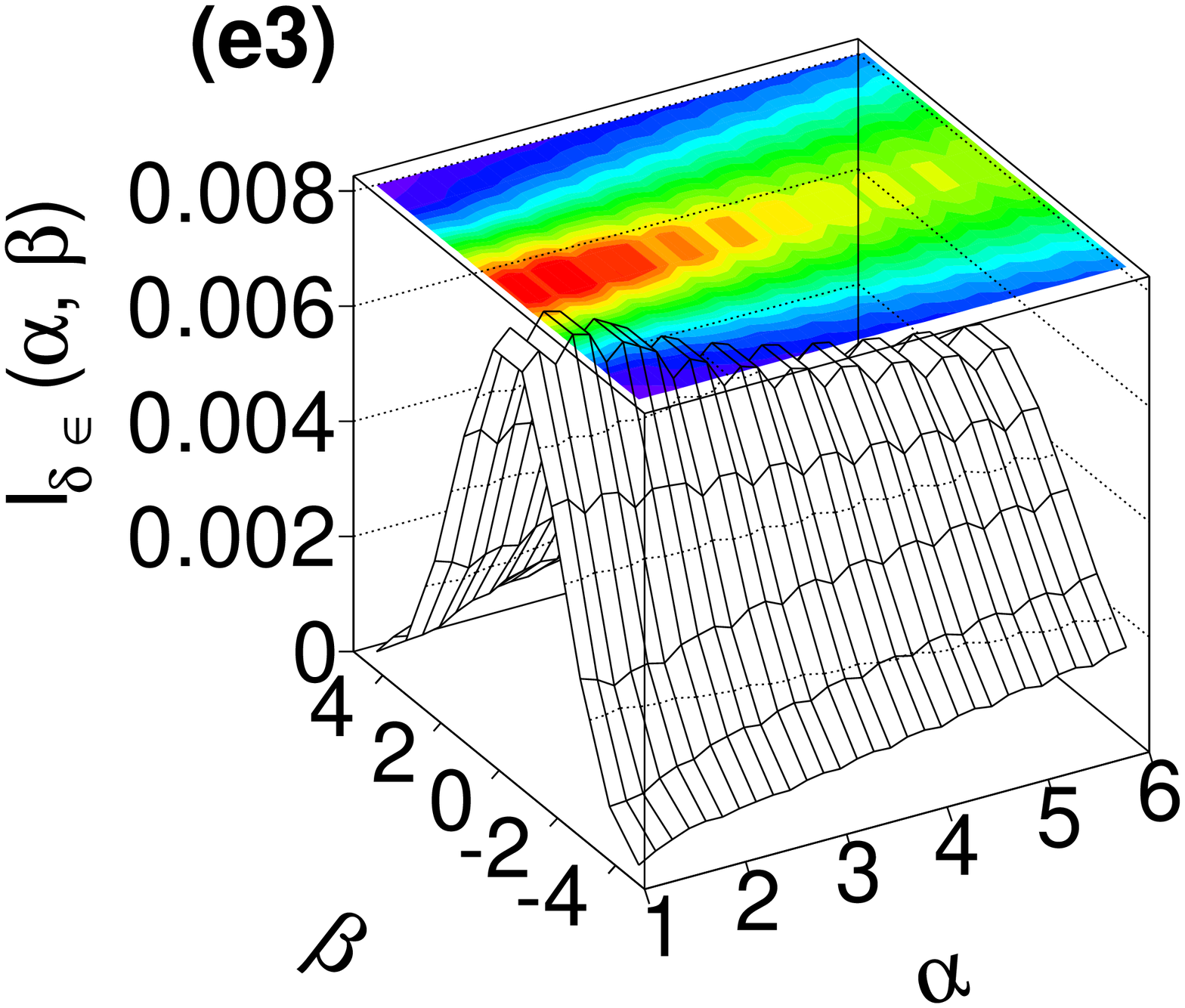}

\caption{(Color on line) Three-dimensional plots and the corresponding
  contour plots for (a) $I_{g_{Tz}}$, (b) $I_{g_{Ty}}$, (c)
  $I_{g_{Lz}}$, (d) $I_{g_{Ly}}$ and (e) $I_{g_{\delta\epsilon}}$ defined in
  Eqs.~(\ref{modgtz}),~(\ref{modgty}),~(\ref{modglz}),~(\ref{modgly}), 
  and~(\ref{moddeltaeps}) as functions of $\alpha$ and $\beta$. The
  plots are shown from a minimum values of
  (1) $\alpha_{\mbox{\tiny{min}}}=0.1$ 
  , (2) $\alpha_{\mbox{\tiny{min}}}=0.5$ and
  (3) $\alpha_{\mbox{\tiny{min}}}=1$ up to a maximum value of
  $\alpha_{\mbox{\tiny{max}}}=6$.}
\label{fig4}
\end{center}
\end{figure*}


where $\phi$ is the angle that the momentum vector ${\mathbf{p}}$
makes with the $\hat{z}$ axis. We use Bjorken's geometry; thus,
\bea
   d^3p&=&p_Tdp_Td\phi dp_x\nn,\\
   p_x&=&p_T\sinh y\nn,\\
   dp_x&=&p_T\cosh y\  dy\nn,\\
   E&=&p_T\cosh y,
   \label{dist}
\eea
 and therefore
\bea
    E\frac{dN}{d^3p}=\frac{1}{p_Tdp_T}\frac{dN}{d\phi dy}.
    \label{dist2}
\eea
For simplicity we consider a freeze-out hypersurface of constant time,
\bea
   d\Sigma_\mu=(d^3r,0,0,0).
   \label{hypersurface}
\eea
Therefore, using Eqs.~(\ref{Cooper-Frye}) and~(\ref{dist2}), the
particle azimuthal distribution around the direction of motion of a
fast moving parton within the $p_T$ interval $p_T^{\mbox{\tiny{min}}}
\leq p_T \leq p_T^{\mbox{\tiny{miax}}}$ is given by
\bea
   \left.\frac{dN}{dy}\right|_{y\simeq0}&=&\frac{\Delta\tau\Delta y}{(2\pi )^3}
   \int_{p_T^{\mbox{\tiny{min}}}}^{p_T^{\mbox{\tiny{max}}}}
   dp_Tp_T^2\nn\\
   &\times&\int d^2r [f(p\cdot u) - f(p_0)],
   \label{distapprox}
\eea
with $\Delta\tau$ the freeze-out time interval and where we have
assumed a perfect correlation between the space-time rapidity $\eta$ and
$y$ to substitute $\Delta\eta$ by $\Delta y$. We assume that the
equilibrium distribution is of the Boltzmann type. In this way, we have
\bea
   f(p_0)&=&\exp[p_T/T_0]\nn\\
   f(p\cdot u)&=&\exp\left[\frac{p_T}{(T_0+\delta T)}
   \left(1-\frac{{\mathbf{g}}_y\sin\phi + {\mathbf{g}}_z\cos\phi}{\epsilon_0(1+c_s^2)}\right)\right],\nn\\
   \label{distexpl}
\eea
where $T_0$ is the background medium's temperature and $\delta T$ is
the change in temperature caused by the passing of the fast
parton. Assuming that the energy density and temperature are related
through Boltzmann's law
\bea
   \epsilon\propto T^4,
   \label{bolaw}
\eea
one gets
\bea
   \frac{\delta T}{T_0}=\frac{\delta\epsilon}{4\epsilon_0}.
   \label{delbolaw}
\eea
Since for the validity of linearized hydrodynamics, both
$\delta\epsilon$ and ${\mathbf{g}}$ need to be small quantities
compared to $\epsilon_0$, we can expand the difference $f(p\cdot u) -
f(p_0)$ to linear order. Using Eqs.~(\ref{distexpl})
and~(\ref{delbolaw}) we get
\bea 
f(p\cdot u) - f(p_0)&\simeq&
\left(\frac{p_T}{T_0}\right)\exp\left[ -p_T/T_0
  \right]\nn\\ &\times&\left( \frac{\delta\epsilon}{4\epsilon_0} +
\frac{{\mathbf{g}}_y\sin\phi + {\mathbf{g}}_z\cos\phi}{\epsilon_0(1+c_s^2)} \right),
   \label{linearorder}
\eea
Therefore, the particle azimuthal distribution around the direction of
motion of a fast moving parton is given, in the linear approximation,
by
\bea
   \left.\frac{dN}{dy}\right|_{y\simeq0}&=&\frac{\Delta\tau\Delta y}{(2\pi )^3}
   \int_{p_T^{\mbox{\tiny{min}}}}^{p_T^{\mbox{\tiny{max}}}}
   dp_T\frac{p_T^3}{T_0}\exp\left[ -p_T/T_0 \right]\nn\\
   &\times&\int d^2r
   \left( \frac{\delta\epsilon}{4\epsilon_0} + \frac{{\mathbf{g}}_y\sin\phi + {\mathbf{g}}_z\cos\phi}{\epsilon_0(1+c_s^2)} \right).
   \label{lineardist}
\eea
Notice that the shape of the distribution around the direction of the
fast moving parton depends on the quantities
\bea
   G_y&\equiv&\int d^2r {\mathbf{g}}_y\nn,\\
   G_z&\equiv&\int d^2r {\mathbf{g}}_z.
   \label{momintegrated}
\eea
When $G_y>G_z$ the distribution is dominated by the $\sin\phi$ factor,
giving rise to two peaks away from $\phi=0$. However, when $G_z>G_y$
the opposite happens and the distribution is dominated by the
$\cos\phi$ factor with the two peaks close to $\phi=0$. These peaks
become a single peak in the extreme case where $G_z\gg G_y$. This is
illustrated in Fig.~\ref{fig3}. 

We now proceed to show that when the velocity of the moving parton is
larger than the speed of sound, $G_y$ is mostly made out of
longitudinal (sound) modes, whereas $G_z$ is mostly made out of wake
(transverse) modes and that the latter dominates the former, giving
rise to a particle distribution centered around the direction of
motion of the moving parton.

\section{Linear viscous hydrodynamics}\label{III}

To compute the $g_y$ and $g_z$ components of the momentum
density vector ${\mathbf{g}}$ which is deposited into the medium by
the fast moving parton, we resort to using linearized viscous
hydrodynamics. Assuming that the disturbance introduced by the parton
is small, the medium's energy momentum tensor can be written as
\bea
   \Theta^{\mu\nu} = \Theta_0^{\mu\nu} + \delta \Theta^{\mu\nu},
   \label{tensor} 
\eea
where $\delta \Theta^{\mu\nu}$ is the perturbation generated by the
parton and $\Theta_0^{\mu\nu}$ is the equilibrium energy momentum
tensor of the underlying medium. Therefore, each of these components
satisfies the equations
\bea
   \partial_\mu \delta \Theta^{\mu\nu} &=& J^\nu\nn,\\
   \partial_\mu \Theta_0^{\mu \nu} &=& 0,
\label{lin_source}
\eea
where $J^\nu$ represents the source of the disturbance, which in this
case corresponds to the fast moving parton. Here we consider that the
parton can be represented by a localized disturbance of the form
\bea 
J^\nu({\mathbf{x}},t)=\left(\frac{dE}{dx}\right) v^\nu\delta
({\mathbf{x}}-{\mathbf{v}}t),
\label{deltasource}
\eea
where $\left( dE/dx\right)$ is the energy loss per unit length and
\bea 
v^\nu= (1,{\mathbf{v}}), 
\label{velsource}
\eea

Effects of a finite source
structure were studied in Ref.~\cite{Neufeld1}, where it is found
that differences in the energy density deposition between a finite
extent source and a localized one exist only close to the
source. Because a hydrodynamic description is anyway valid only for
large distances from the source, compared to the transport mean free
path, here we consider that the description of a source as
localized exactly at the position of the fast parton; suffices.

Furthermore, we consider that the partons produced 
in the hard scattering are propagating asymptotically, 
that is, we ignore any finite time effects associated with 
the initial hard scattering during the collision. 
The latte have been raised and examined 
in a different context for example in Ref.~\cite{Gossiaux}. 
Although these effects are important and may very well be significant for the description of the energy loss of a fast moving parton in finite size media in this work our focus is on the time dependence associated with energy momentum deposited after the hard partons are produced, that is, while traveling into the medium, and on how the shape of the away side can be understood in terms of the energy deposited by two partons instead of one parton.

To describe the propagation of the disturbance caused by the source, Eqs. (\ref{lin_source}) 
are solved by considering that the energy momentum tensor $\Theta^{\mu\nu}$ consists
of a piece that describes an isotropic fluid,
\bea
\Theta_0^{\mu \nu} = -p g^{\mu\nu} + (\epsilon + p)u_0^\mu u_0^\nu
\eea
and a disturbance $\delta\Theta^{\mu\nu}$ caused by the source, whose explicit components
to first order in shear ($\eta$) and bulk ($\zeta$) viscosity are given by 
\bea
\label{thetapert}
\delta\Theta^{00} &=& \delta \epsilon, \nonumber\\
\delta\Theta^{0i} &=& {\mathbf g}, \nonumber \\
\delta\Theta^{ij} &=& \delta_{ij}c_s^2 \delta\epsilon - \frac{3}{4}\Gamma_s(\partial^i\mathbf{g}^j + \partial^j\mathbf{g}^i -\frac{2}{3}\delta_{ij}\nabla \cdot {\mathbf g} )  \nonumber \\
&-&\zeta ~\delta_{ij}\nabla \cdot {\mathbf g},
\eea
where $\epsilon (t, \mathbf{x}) = \epsilon_0 + \delta \epsilon (t,\mathbf{x})$
with $\epsilon_0$ correspond to the energy density of the background fluid,
$\delta\epsilon$ corresponds to the energy density associated to the disturbance, and
\bea
\Gamma_s\equiv \frac{4 \eta }{3\epsilon_0(1+c_s^2)}
\label{defGammas}
\eea
is the sound attenuation length.

In the linear approximation and vanishing bulk viscosity~\cite{Neufeld1}, 
the dynamic description of the propagation of the disturbance is given by
the first of Eqs.~(\ref{lin_source}), whose explicit components can be written
as
\bea
\label{beforefourier}
\partial_0 \delta\epsilon + \nabla\cdot\mathbf{g} &=& J^0, \nonumber \\
\partial_0 \bd{g}^i + \partial_j \delta\Theta^{ij} &=& J^i.
\eea 
These equations can be readily solved in momentum space. We define 
the Fourier transform pair $f({\mathbf{x}},t)$ and
$f({\mathbf{k}},\omega)$ as
\bea 
f({\mathbf{x}},t) = \frac{1}{(2 \pi)^4}\int d^3 k \int d \omega
\,e^{i \bd{k}\cdot\bd{x} - i \omega t} f({\mathbf{k}},\omega).
\label{FT}
\eea

Using Eq.~(\ref{FT}) in Eqs.~(\ref{beforefourier}), together with Eqs.~(\ref{thetapert}),
we obtain
\bea
\label{afterfourier}
-i\omega \delta\epsilon + i \mathbf{k}\cdot\mathbf{g} &=& J^0, \nonumber \\
-i\omega\mathbf{g}^i + ic_s^2k^i\delta\epsilon + \frac{3}{4}\Gamma_s[k^2\mathbf{g}^i 
+ \frac{k^i}{3}(\mathbf{k}\cdot\mathbf{g})] &=& J^i.
\eea 
Thus the perturbed energy density, $\delta
\epsilon$ and momentum density components $\bd{g}^i$, can be obtained by solving
the algebraic system of Eqs.~(\ref{afterfourier}), and the solutions are given by~\cite{Neufeld2}
\bea 
   \delta\epsilon
   ({\mathbf{k}},\omega) &=& \frac{i {\mathbf{k}}\cdot{\mathbf{J}}({\mathbf{k}},\omega) +
   J^0({\mathbf{k}},\omega)(i \omega - \Gamma_s k^2)}{\omega^2 - c_s^2
   k^2 + i \Gamma_s \omega k^2}, 
\label{eps}\\
   {\mathbf{g}}_L ({\mathbf{k}},\omega) &=& \frac{i\left[\frac{\omega}{k^2}
   {\mathbf{k}}\cdot{\mathbf{J}}({\mathbf{k}},\omega)+ c_s^2 J^0({\mathbf{k}},\omega)\right]{\mathbf{k}}}
   {\omega^2 - c_s^2 k^2 + i \Gamma_s \omega k^2}, 
\label{gl}\\
    \bd{g}_T({\mathbf k},\omega) &=& \bd{g} - \bd{g}_L = \frac{i{\mathbf
    J}_T({\mathbf k},\omega)}{\omega + i \frac{3}{4}\Gamma_s k^2},
\label{gt}
\eea
where we have written the momentum density vector in terms of its
transverse and longitudinal components with respect to the Fourier
mode $\mathbf{{k}}$, namely,
\bea
   {\mathbf{g}} = {\mathbf{g}}_L + {\mathbf{g}}_T,
\label{translong}
\eea
with the definition of longitudinal and transverse components of any
vector $\mbox{\boldmath${\sigma}$}$ given by
\bea
   \mbox{\boldmath${\sigma}$}_L&\equiv&\frac{(\mbox{\boldmath${\sigma}$}
   \cdot{\mathbf{k}})}{k^2}{\mathbf{k}}\nn,\\
   \mbox{\boldmath${\sigma}$}_T&\equiv&\mbox{\boldmath${\sigma}$} - \mbox{\boldmath${\sigma}$}_L.
\label{expltranslong}
\eea
Note that the source term $J^\nu$ in Eq.~(\ref{deltasource}) is Fourier transformed to
\bea 
J^\nu(\bd{k}, \omega) = (2\pi) \left(\frac{d E}{d
  x}\right)\delta(\bd{k\cdot v} - \omega)v^\nu,   
\label{simpleJk}
\eea
where we assumede a constant energy loss per unit length.

From Eqs.~(\ref{eps})--(\ref{gt}) one can obtain the space-time
solutions for $\delta\epsilon({\mathbf{x}},t)$ and
${\mathbf{g}}({\mathbf{x}},t)$ upon use of Eq.~(\ref{FT}). The
corresponding expressions are
\bea 
\bd{g}_T(\bd{x}, t) &=& i  (2\pi) \left( \frac{d E}{d x}
\right) \int \frac{d^3 k}{(2\pi )^3}\int \frac{d \omega }{(2\pi )}e^{i
  \bd{k}\cdot\bd{x} - i \omega t}\nn\\ &\times& \co{\bd{v} -
  \frac{(\bd{k\cdot J}) \bd{k}} {k^2}}\frac{\delta(\bd{k}\cdot\bd{v} -
  \omega)}{\omega + i \frac{3}{4} \Gamma_s k^2} 
\label{gT(x)}
\eea
\bea 
\bd{g}_L(\bd{x}, t) &=& i  (2\pi) \left( \frac{d E}{d x}
\right) \int \frac{d^3 k}{(2\pi )^3}\int \frac{d \omega }{(2\pi )}e^{i
  \bd{k}\cdot\bd{x} - i \omega t}\nn\\ &\times& \bd{k}
\co{\frac{\omega}{k^2}\bd{k}\cdot\bd{v} +
  c_s^2}\frac{\delta(\bd{k}\cdot\bd{v} - \omega)}{\omega^2 - c_s^2k^2
  + i \Gamma_s \omega k^2} 
\label{gL(x)}
\eea
and
\bea 
\delta\epsilon (\bd{x}, t) &=& i (2\pi) \left( \frac{d
  E}{d x} \right) \int \frac{d^3 k}{(2\pi )^3}\int \frac{d \omega
}{(2\pi )}e^{i \bd{k}\cdot\bd{x} - i \omega t}\nn\\ &\times&\co{ i
  \bd{k}\cdot\bd{v} + i \omega - \Gamma_s
  k^2}\frac{\delta(\bd{k}\cdot\bd{v} - \omega)}{\omega^2 - c_s^2k^2 +
  i \Gamma_s \omega k^2}.\nn\\ 
\label{deltaepsilon(x)}
\eea
To compute the integrals in
Eqs.~(\ref{gT(x)})--(\ref{deltaepsilon(x)}) we use cylindrical
coordinates with the $k_z$ direction along the direction of motion,
${\mathbf{v}}$, of the fast parton. It is easier to start with the
components of $\bd{g}_T$. After carrying out the frequency and angular
integration, we get for the $z$ component
\bea 
\left(\bd{g}_T\right)_z &=& i (2\pi) \left( \frac{d E}{d
  x} \right) v \int_0^\infty \frac{dk_T}{(2\pi
  )^2}\int_{-\infty}^\infty \frac{dk_z}{(2\pi )}e^{i
  k_z(z-vt)}\nn\\ &\times& \frac{k_T^3{\mbox{J}}_0(k_Tx_T)} {(k_z^2
  + k_T^2)[vk_z + i \frac{3}{4} \Gamma_s(k_z^2 + k_T^2) ]}, 
   \label{gT(x)expl} 
\eea

where ${\mbox{J}}_0$ is a Bessel function and $x_T=\sqrt{y^2}$ is the distance
from the parton along the transverse direction (along the $\hat{y}$ axis
in the geometry we are using). The
integration over $k_z$ is performed using contour integration. For
causal motion ($z-vt >0$) we close the contour on the lower half
$k_z$-plane. The poles that contribute are located at
\bea 
k_z=\left\{
   \begin{array}{c}
   -ik_T,\\
   i\frac{2v}{3\Gamma_s}\left(1-\sqrt{1+\left( \frac{3\Gamma_sk_T}{2v} \right)^2}\right).
   \end{array}\right.
\label{poles}
\eea

After carrying out the $k_z$ integration, the remaining integral can
be expressed in terms of the dimensionless quantities
\bea
   \xi&\equiv&\left(\frac{3\Gamma_s}{2v}\right)k_T\nn,\\
   \alpha&\equiv&|z-vt|/\left(\frac{3\Gamma_s}{2v}\right)\nn,\\
   \beta&\equiv&x_T/\left(\frac{3\Gamma_s}{2v}\right),
\label{dimensionless}
\eea
as
\bea 
\left(\bd{g}_T\right)_z &=& -\left(\frac{1}{4\pi}\right)
\left( \frac{d E}{d x}
\right)\left(\frac{2v}{3\Gamma_s}\right)^2\int_0^\infty d\xi \ \xi
      {\mbox{J}}_0(\xi \beta )\nn\\ &\times& \left[e^{-\alpha\xi} +
        \frac{e^{-\alpha(\sqrt{1+\xi^2}-1)}}{[\sqrt{1+\xi^2}-(1+\xi^2)]}\right]. 
\label{remaininggtz}
\eea
Notice that although the source term describes infinite propagation, in practice the initial and final times of the evolution are implemented by considering a finite interval for the variable $\alpha$, which represents the distance to the source in units of the sound attenuation length.

The first term in Eq.~(\ref{remaininggtz}) can be analytically
integrated. For numerical purposes, it is more convenient to rewrite
the second term after the change of variable:
\bea
   s=\sqrt{1+\xi^2}-1.
\label{change}
\eea
The final result is
\bea
   \left(\bd{g}_T\right)_z &=& \left(\frac{1}{4\pi}\right)
   \left( \frac{d E}{d x} \right)\left(\frac{2v}{3\Gamma_s}\right)^2
   \left[-\frac{\alpha}{\pa{\alpha^2+\beta^2}^{\frac{3}{2}}}\right.\nn\\
   &+&\left.\int_0^{\infty}ds(s
   +2){\mbox{J}}_0\pa{\beta\sqrt{s(s+2)}}e^{-\alpha
   s}\right]\nn\\ 
   &\equiv&
   \left(\frac{1}{4\pi}\right)
   \left( \frac{d E}{d x} \right)\left(\frac{2v}{3\Gamma_s}\right)^2 I_{g_{Tz}}(\alpha ,\beta ).
\label{modgtz}
\eea
In a similar fashion we get
\bea
   \left(\bd{g}_T\right)_y &=& \left(\frac{1}{4\pi}\right)
   \left( \frac{d E}{d x} \right)\left(\frac{2v}{3\Gamma_s}\right)^2
   \left[\frac{\beta}{\pa{\alpha^2+\beta^2}^{\frac{3}{2}}}\right.\nn\\
   &-&\left.\int_0^{\infty}ds\sqrt{s(s+2)}{\mbox{J}}_1\pa{\beta\sqrt{s(s+2)}}e^{-\alpha
   s}\right]\nn\\
   &\equiv&
   \left(\frac{1}{4\pi}\right)
   \left( \frac{d E}{d x} \right)\left(\frac{2v}{3\Gamma_s}\right)^2 I_{g_{Ty}}(\alpha ,\beta ). 
\label{modgty}
\eea

The computation of the components of ${\mathbf{g}}_L$ is more involved
due to the analytic structure of the poles in the complex
$k_z$-plane. Let us fist compute the $z$ component. After carrying out
the frequency and angular integration we get
\bea
   \left(\bd{g}_L\right)_z &=& i (2\pi) \left( \frac{d E}{d x} \right)
   \int_0^\infty \frac{dk_T}{(2\pi )^2}\int_{-\infty}^\infty \frac{dk_z}{(2\pi )}e^{i k_z(z-vt)}\nn\\
   &\times&
   \frac{k_T{\mbox{J}}_0(k_Tx_T)k_z[k_z^2+\frac{c_s^2}{v^2}(k_z^2+k_T^2)]} 
   {(k_z^2+k_T^2)[k_z^2 + (i \frac{\Gamma_s}{v}k_z - \frac{c_s^2}{v^2})(k_z^2 + k_T^2) ]}.
   \label{gL(x)expl}
\eea

Notice that the integrand contains the parameter $c_s^2/v^2$ and that for a fast moving parton $v\simeq 1$. 

This allows for an approximation to be implemented in order to render 
the results more transparent: We consider the parameter $c_s^2/v^2<1$ to be small such that we can 
expand the integrands before calculating the remaining integral.  This approximation is sustained by lattice estimates (see, for instance, Ref.~\cite{Borsanyi}) of the speed of sound which 
show that $c_s$ increases monotonically from about one-third 
of the ideal gas limit ($\simeq \sqrt{1/3}$) for $T \gtrsim 1.5 T_c$ and approaches this limit only for $T>4T_c$, where $T_c$ is the critical temperature for the phase transition. Therefore, even though the approximation $c_s^2/v^2<1$ could only be regarded as a parametric limit, the fact that for experimental conditions $c_s^2 < 1/3$ makes 
the approximation to be a good one for the present context. Here we use $c_s^2=1/3$, which can then be taken 
as a worst-case scenario for a numerical estimate. 
Thus, for conditions close to the ones
after a heavy-ion reaction $c_s^2/v^2<1$ and we can expand the
integrand in Eq.~(\ref{gL(x)expl}) in this 
parameter. Notice that in the literature there are different approximation schemes to carry out the integrals such as the ones in Eq.~(\ref{gL(x)expl}). For instance, in order to calculate the energy momentum deposited by 
longitudinal modes, the approximation $s^2/\Gamma_s \gg k_T$ 
is made in Refs.~\cite{Neufeld3} and~\cite{Neufeld4}, which allows to integrate over $k_z$ analytically, before proceeding to the numerical computation of the integral over $k_T$. 

Therefore, to first order in $c_s^2/v^2$, we get
\bea
   \left(\bd{g}_L\right)_z &=& i (2\pi) \left( \frac{d E}{d x} \right)
   \int_0^\infty \frac{dk_T}{(2\pi )^2}k_T{\mbox{J}}_0(k_Tx_T)\nn\\
   &&
    \int_{-\infty}^\infty \frac{dk_z}{(2\pi )}\nn\\
   &\times&
   \left\{
   \frac{k_z^2e^{i k_z(z-vt)}} 
   {(k_z^2+k_T^2)[k_z + (i \frac{\Gamma_s}{v})(k_z^2 + k_T^2) ]}\right.\nn\\
   &+&
   \left.
   \left(\frac{c_s^2}{v^2}\right)
   \frac{2k_z(i \frac{\Gamma_s}{v})(k_z^2 + k_T^2)e^{i k_z(z-vt)}}
    {[k_z + (i \frac{\Gamma_s}{v})(k_z^2 + k_T^2) ]}
   \right\}.\nn\\ 
   \label{gL(x)expl-2}
\eea
For causal motion ($z-vt >0$) we close the contour on the lower half
$k_z$-plane. The poles that contribute are located at
\bea
   k_z=\left\{
   \begin{array}{c}
   -ik_T,\\
   i\frac{v}{2\Gamma_s}\left(1-\sqrt{1+\left( \frac{2\Gamma_sk_T}{v} \right)^2}\right).
   \end{array}\right.
\label{poles-2}
\eea

The remaining integral over $k_T$ can be expressed in terms of the
dimensionless quantities defined in Eq.~(\ref{dimensionless}). For
numerical purposes, it is more convenient to rewrite this integral after
 the change of variable in Eq.~(\ref{change}). The result
reads as follows:
\bea
   \left(\bd{g}_L\right)_z &=& \left(\frac{1}{4\pi}\right)
   \left( \frac{d E}{d x} \right)\left(\frac{2v}{3\Gamma_s}\right)^2
   \left[
   \frac{\alpha}{\pa{\alpha^2+\beta^2}^{\frac{3}{2}}}\right.\nn\\
   &-&
   \frac{9}{16}\int_0^{\infty}ds   
   {\mbox{J}}_0\left(\frac{3}{4}\beta\sqrt{s(s+2)}\right)
   e^{-\frac{3}{4}\alpha s}\left(s\right.\nn\\
   &+&
   \left.\left.c_s^2/v^2\left(
   \frac{3\alpha s}{2(s+1)}-\frac{2}{(s+1)^2}-2\right)\right)\right]\nn\\
   &\equiv&
   \left(\frac{1}{4\pi}\right)
   \left( \frac{d E}{d x} \right)\left(\frac{2v}{3\Gamma_s}\right)^2 I_{g_{Lz}}(\alpha ,\beta ).
\label{modglz}
\eea
In a similar fashion we obtain
\bea
   \left(\bd{g}_L\right)_y &=& \left(\frac{1}{4\pi}\right)
   \left( \frac{d E}{d x} \right)\left(\frac{2v}{3\Gamma_s}\right)^2
   \left[
   \frac{\beta }{\pa{\alpha^2+\beta^2}^{\frac{3}{2}}}\right.\nn\\
   &+&
   \frac{9}{16}\int_0^{\infty}ds   
   {\mbox{J}}_1\left(\frac{3}{4}\beta \sqrt{s(s+2)}\right)
   e^{-\frac{3}{4}\alpha s}\left(-1\right.\nn\\
   &+&
   \left.\left.
   c_s^2/v^2\pa{\frac{2}{(s+1)^2}-\frac{3\alpha}{2(s+1)}-\frac{2}{s}}\right)\right]\nn\\
   &\equiv&
   \left(\frac{1}{4\pi}\right)
   \left( \frac{d E}{d x} \right)\left(\frac{2v}{3\Gamma_s}\right)^2 I_{g_{Ly}}(\alpha ,\beta ),
\label{modgly}
\eea
and
\bea
   \delta\epsilon&=&\left(\frac{1}{4\pi}\right)
   \left( \frac{d E}{d x} \right)\left(\frac{2v}{3\Gamma_s}\right)^2\left(\frac{9}{8v}\right)\nn\\
   &\times&
   \int_0^{\infty}ds{\mbox{J}}_0\left(\frac{3}{4}\beta \sqrt{s(s+2)}\right)
   e^{-\frac{3}{4}\alpha s}\nn\\
   &\times&
   \co{1+\frac{2c_s^2/v^2}{s}\pa{-1+\frac{1+s\pa{3+s+\frac{3}{4}\alpha(s+1)}}{(s+1)^2}}}\nn\\
   &\equiv&
   \left(\frac{1}{4\pi}\right)
   \left( \frac{d E}{d x} \right)\left(\frac{2v}{3\Gamma_s}\right)^2 
   \left(\frac{9}{8v}\right)I_{\delta\epsilon}(\alpha ,\beta ).
\label{moddeltaeps}
\eea

\begin{figure}[h] 
\begin{center}
\includegraphics[scale=0.4]{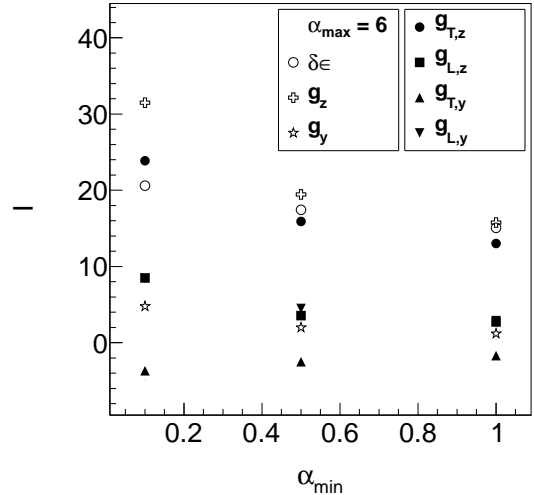}
\caption{Integrals of the functions $I_{g_{Tz}}, I_{g_{Ty}},
  I_{g_{Lz}}, I_{g_{Ly}}$ and $I_{g_{\delta\epsilon}}$ over the domain
  $\alpha_{\mbox{\tiny{min}}}<\alpha<\alpha_{\mbox{\tiny{max}}}$,
  $-5<\beta<5$ for the different values of
  $\alpha_{\mbox{\tiny{min}}}$. We also plot the values of the
  combinations $I_z\equiv I_{g_{Tz}} + I_{g_{Lz}}$ and $I_y\equiv
  I_{g_{Ty}} + I_{g_{Ly}}$. Notice that, for all of values of
  $\alpha_{\mbox{\tiny{min}}}$, the largest contribution to $I_z$
  comes from $I_{g_{Tz}}$, which is the wake or transverse mode, whereas
  the largest contribution to $I_y$ comes from $I_{g_{Ly}}$, which is
  the sound or longitudinal mode. Moreover, $I_z$ is always larger
  than $I_y$.}
\label{fig5}
\end{center}
\end{figure}

Figure~\ref{fig4} shows three-dimensional plots of $I_{g_{Tz}},
I_{g_{Ty}}, I_{g_{Lz}}, I_{g_{Ly}}$ and $I_{g_{\delta\epsilon}}$ as
functions of $\alpha$ and $\beta$. Shown also are the corresponding
contour plots. To test the sensitivity of the results on the distance
to the source, the plots are shown starting from a minimum value of
$\alpha_{\mbox{\tiny{min}}}=0.1,0.5,1$ up to a maximum value of
$\alpha_{\mbox{\tiny{max}}}=6$. Figure~\ref{fig5} shows the integrals
of the above functions over the domain
$\alpha_{\mbox{\tiny{min}}}<\alpha<\alpha_{\mbox{\tiny{max}}}$,
$-5<\beta<5$ for the different values of
$\alpha_{\mbox{\tiny{min}}}$. Shown also are the values of the
combinations $I_z\equiv I_{g_{Tz}} + I_{g_{Lz}}$ and $I_y\equiv
I_{g_{Ty}} + I_{g_{Ly}}$. Notice that, for all of the shown values of
$\alpha_{\mbox{\tiny{min}}}$, the largest contribution to $I_z$ comes
from $I_{g_{Tz}}$, which is the wake or transverse mode, whereas the
largest contribution to $I_y$ comes from $I_{g_{Ly}}$, which is the
sound or longitudinal mode. Moreover, $I_z$ is always larger than
$I_y$. This last result shows that for the case treated here, where
$c_s^2/v^2 < 1$, the momentum deposition is preferentially
forward peaked.

\clearpage

\section{Azimuthal correlations}\label{IV}

\begin{figure}[h] 
\begin{center}
\includegraphics[scale=0.5]{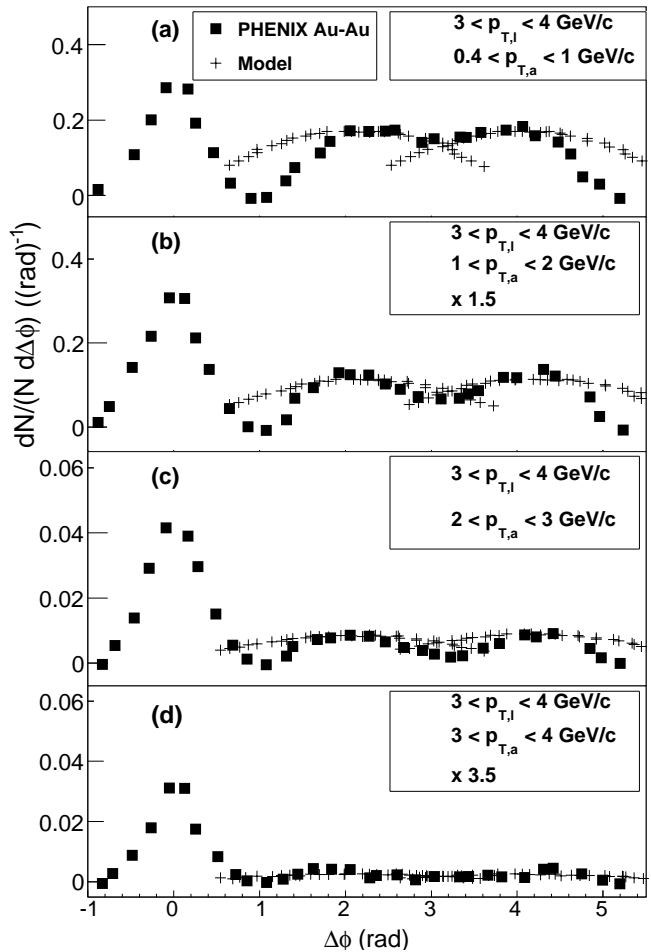}
\caption{Per-trigger azimuthal angular correlations. Shown are four
  different cases, each one having the leading hadron produced with a
  fixed momentum range 3 GeV $\leq p_T \leq$ 4 GeV. The associated
  hadrons are in following the momentum ranges: (a) 0.4 GeV $\leq p_T
  \leq$ 1 GeV, (b) 1 GeV $\leq p_T \leq$ 2 GeV, (c) 2 GeV $\leq p_T
  \leq$ 3 GeV and (d) 3 GeV $\leq p_T \leq$ 4 GeV. The theoretical
  distributions are multiplied by the same correction factor,
  $f\approx 30$. The results are compared to data from
  PHENIX~\cite{PHENIX}.}
\label{fig6}
\end{center}
\end{figure}

\begin{figure}[htb!]
\begin{center}
\includegraphics[scale=0.4]{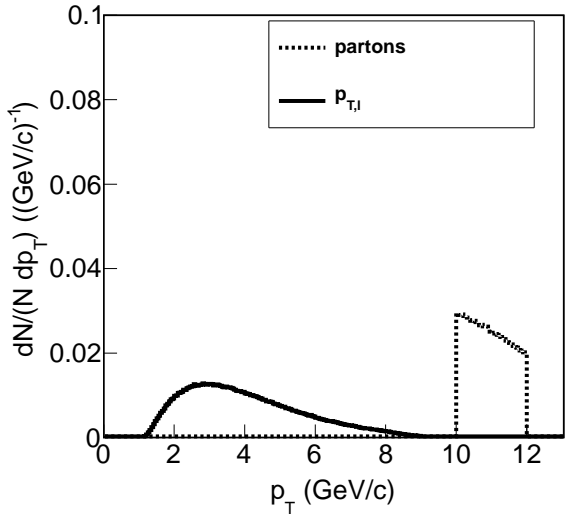}
\caption{Hadron momentum distribution obtained from a leading
parton with momentum in the range 10 GeV $\leq p_T \leq$ 12 GeV.}
\label{fig7}
\end{center}
\end{figure}

\begin{figure}[t]
\begin{center}
\includegraphics[scale=0.4]{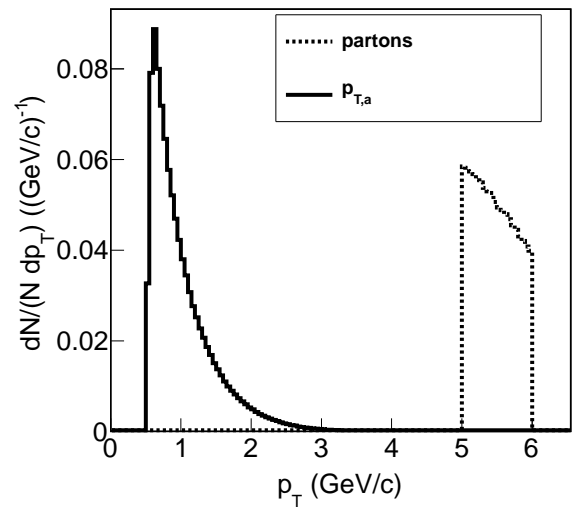}
\caption{Hadron momentum distribution obtained from an associated
parton with momentum in the range 5 GeV $\leq p_T \leq$ 6 GeV.}
\label{fig8}
\end{center}
\end{figure}

We now proceed to use the formalism developed above to compute the
azimuthal angular correlations for events where the leading hadron has
a momentum larger than or equal to the associated
ones. Figure~\ref{fig6} shows the per-trigger correlations for the
cases where the leading hadron is produced in a momentum range 3 GeV
$\leq p_T \leq$ 4 GeV and the associated ones in the momentum ranges
0.4 GeV $\leq p_T \leq$ 1 GeV, 1 GeV $\leq p_T \leq$ 2 GeV, 2 GeV
$\leq p_T \leq$ 3 GeV and 3 GeV $\leq p_T \leq$ 4 GeV, compared to
data from PHENIX~\cite{PHENIX}. To obtain these correlations we have
generated $2 \to 3$ hard scattering parton events using the MadGraph5
event generator~\cite{madgraph}, within the momentum window 10 GeV
$\leq p_T \leq$ 12 GeV.  Out of the three partons, we choose the one
with the largest momentum to become the leading hadron by collinear
fragmentation using the Kniehl, Kramer and Pötter (KKP) parton
fragmentation functions~\cite{KKP}. The other two partons in the event
travel within the medium and thus represent the sources of energy
momentum deposition in the away side. We consider that these partons
travel 6 fm on average~\cite{Ayala1}, with a medium-induced energy
loss per unit length $dE/dx =$ 2 GeV/fm. Therefore, on average, these
partons transfer all of their initial momentum to the medium. For
these away side partons, we compute the angular distribution around
their direction of motion by means of the Cooper-Frye and linear
viscous hydrodynamics formalisms, described in Secs.~\ref{II}
and~\ref{III}. We use the value $\alpha_{\mbox{\tiny{min}}}=0.1$.
Since we consider massless partons, we take their speed to be
$v = 0.9995c$. The sound attenuation length is taken to be $\Gamma_s =
1/(3 \pi T_0)$, which results from taking $\eta = 1/(4\pi)$ and
considering the value $T_0 = 350$ MeV, and the sound velocity is $c_s =
1/\sqrt{3}$. To produce final-state hadrons out of the away
side partons that emerge via the Cooper-Frye formula, we consider that
the partons in the head shock region carry half of the hard-scattering
parton momentum. We subsequently let these partons produce final-state
hadrons using once again the KKP in vacuum fragmentation
functions. This procedure is intended to mimic surface emission of the
leading parton together with in-medium energy momentum deposited by
the away side ones. The distributions thus obtained are multiplied by
the same correction factor, $f\approx 30$. This factor is introduced
for data comparison purposes and has to be regarded as a quantitative
measure of the dynamical details left out in this analysis. However, we
emphasize that this factor is the same for all bins shown. The hadron
momentum distributions obtained from the leading and associate partons
are shown in Figs.~\ref{fig7} and~\ref{fig8}, respectively.

Notice that Fig.~\ref{fig6} shows a quite good agreement of this
simple scenario with the correlation data, particularly because it
reproduces the systematics of double-hump decreasing intensity as the
momentum of the away side hadrons becomes closer to the momentum of
the leading hadron.

\section{Summary and conclusions}\label{V}

In this work, we have studied the way that fast moving partons deposit
their energy momentum when traveling in a medium. We showed that,
for conditions resembling a medium produced in a heavy-ion reaction,
energy momentum is preferentially deposited along the head shock and
that, in order to produce a double-hump structure in the away side of
azimuthal correlations, one can consider that two partons, instead of
one, travel towards the away side. We argued that these away side
partons can be originally produced by a $2\to 3$ hard scattering. Even
though $2 \to 3$ parton processes are suppressed with respect to $2\to
2$ ones by an extra power of $\alpha_s$, there is also a kinematic
enhancement that, partially compensates. This is due to the fact that
for a given energy deposited in the away side, the partons from $2\to
3$ processes are produced with lower momentum than the away side
parton in $2\to 2$ processes; thus, the former are more copiously
produced.

We resorted to linear viscous hydrodynamics to compute the medium
response to the fast moving partons and the Cooper-Frye formula to
compute the parton distribution originated from this disturbance in a
static background medium. To mimic a scenario where the leading hadron
comes from surface emission and the away side partons deposit all of
their original energy momentum within the medium, we selected
partons produced in $2 \to 3$ processes, evolving the parton with the
highest energy to become the leading hadron using KKP in vacuum
fragmentation. Also, the head shock region around the direction of
motion of the away side partons, is assumed to carry half of the
original hard-scattering parton momentum. The partons in the
head shock region are then hadronized using also KKP in vacuum
fragmentation.  The comparison to PHENIX data shows that this simple
scenario reproduces the systematics of a decreasing away side
correlation when the momentum of the associated hadrons becomes closer
to the momentum of the leading hadron.  This scenario seems to avoid
the shortcomings of the Mach cone as the origin of the double-hump
structure in the away side.  

Notice that another possibility to have three partons in the final
state is to start from a $2 \to 2$ process followed by a large angle
parton emission. The probability of this process, however, is suppressed
with respect to (direct) $2 \to 3$ processes. To see this, consider a
given total momentum for the two away side partons. If these come from
a single parton in the away side, which then radiates a parton with a
large angle, the radiating parton needs to be produced with a momentum
of the order of twice the momentum of each of the two final away side
partons. This is to say that, in order for the radiated and the
radiating partons to have their momenta add up to the momentum of
the leading parton, the radiating (original) parton needs to have a
momentum of order twice the momentum of what it will have after
radiating. Therefore, we come back to the phase space suppression
argument we have put forward, namely that producing a parton with a
larger momentum has less phase space than producing partons with
lesser momenta. If we now account for the extra power of $\alpha_s$,
these kinds of processes are suppressed with respect to (direct) $2 \to
3$ processes.

We did not attempt to quantitatively address parton splitting in
$2\to 2$ processes. However, recall that splitting is important in two
situations: (1) the soft and collinear regions and (2) for the
evolution of the parton shower that makes up the final-state jet. For
the first case, collinear and soft divergences are taken care of by
the dipole substraction implemented in MadGraph 5 that we use. On the
other hand, real (as opposed to singular and thus virtually corrected)
splitting needs to be accounted for when pursuing a more detailed
analysis of the jet shapes and their possible in-medium
modification. In this work, we adopted the simpler point of view
that the energy momentum is deposited in the forward direction by the
parton that plows through the medium and not by the other (lesser
momentum) partons that this one may split into. In any case, we
believe that splitting should be important for the shape of the
individual jet but not for the double-hump structures in the
away side, unless the splitting results in a large angle emission, in
which case we come back to the discussion above. In this sense,
splitting should not affect the qualitative description we presented.

The scenario we considered corresponds to the analog case of
surface emission in $2 \to 2$ processes, whereby the leading hadron
comes from a parton emitted outwards from the surface and the
away side hadron comes from a parton emitted towards the interior of
the fireball. In the $2 \to 3$ case, we consider the leading parton
emitted radially outwards and the away side partons to travel within
the fireball an average distance such that, for the energy loss per
unit length we have taken, they deposit on average all their energy
within the medium. A more realistic scenario should consider all the
possibilities, that is, partons traveling towards the fireball
produced with larger and lower energies and different directions. Some
of these will not deposit all of their energy within the medium and
could thus hadronize outside by fragmentation. The details of the
calculation can certainly vary but it is clear that these last kinds
of processes add up to the signal, albeit in a way that needs to be
quantified. Since the purpose of the present work was to call
attention for an alternative process to the usual Mach cone or the
initial-state fluctuation scenarios, we did not carry out such
detailed calculations. We are planning on doing so and reporting
elsewhere shortly.

For the case where the background is not static, one could in
principle consider different scenarios. Since the hard scattering
happens at short times, the main effect should be caused by a dilution
of the medium due to longitudinal expansion, given that transverse
expansion sets in at larger times. This dilution in turn can be
accounted for by a decreasing energy density of the background fluid,
resulting in a larger sound attenuation length and therefore a smaller
energy momentum deposition within the medium. The Cooper-Frye formula
could no longer be implemented at a constant-time hypersurface and one
would need to resort to a Bjorken-like scenario whereby freeze-out
happens over a proper time period which can be related to a
temperature interval. To quantitatively test the differences between a
dynamic and a static scenario, we felt it was important first to have
the latter as the reference to build on top of it with other effects,
such as expansion. We are certainly planning to include these effects
and report on the findings shortly after. A more detailed analysis is
currently under way to explore the systematics of away side
correlations for a larger set of conditions and will be reported
elsewhere.

\section*{Acknowledgments}

Support for this work has been received in part from DGAPA-UNAM under
Grant No. PAPIIT-IN103811, CONACyT-M\'exico under Grant No. 128534
and {\it Programa de Intercambio UNAM-UNISON}.

\end{document}